\documentstyle[preprint,eqsecnum,aps]{revtex}

\newtheorem{theorem}{Theorem}
\newtheorem{acknowledgement}[theorem]{Acknowledgement}

\begin{document}
\draft
\title{Solutions of relativistic wave equations in superpositions of Aharonov-Bohm,
magnetic, and electric fields.}
\author{V.G. Bagrov\thanks{%
On leave from Tomsk State University and Tomsk Institute of High Current
Electronics, Russia}, D.M. Gitman\thanks{%
e-mail: gitman@fma.if.usp.br}, and V.B. Tlyachev\thanks{%
Tomsk Institute of High Current Electronics, Russia}}
\address{Instituto de Fisica, Universidade de S\~{a}o Paulo, \\
C.P 66318, 05315-970 S\~{a}o Paulo,\ SP, Brasil}
\date{\today}
\maketitle

\begin{abstract}
We present new exact solutions (in $3+1$ and $2+1$ dimensions) of
relativistic wave equations (Klein-Gordon and Dirac) in external
electromagnetic fields of special form. These fields are combinations of
Aharonov-Bohm solenoid field and some additional electric and magnetic
fields. In particular, as such additional fields, we consider longitudinal
electric and magnetic fields, some crossed fields, and some special
non-uniform fields. The solutions obtained can be useful to study
Aharonov-Bohm effect in the corresponding electromagnetic fields.
\end{abstract}


\section{Introduction}

Aharonov-Bohm (AB) effect \cite{AhaBo59} plays an important role in quantum
theory refining the status of electromagnetic potentials in this theory.
First this effect was discussed in relation to a study of interaction
between a non-relativistic charged particle and an infinitely long and
infinitesimally thin magnetic solenoid field (further AB field) (a similar
effect was discussed earlier by Ehrenberg and Siday \cite{EhrSi49}). It was
discovered that particle wave functions vanish at the solenoid line. In
spite of the fact that the magnetic field vanishes out of the solenoid, the
phase shift in the wave functions is proportional to the corresponding
magnetic flux \cite{WuYa75}. A non-trivial particle scattering by the
solenoid is interpreted as a possibility for quantum particles to ''feel''
potentials of the corresponding electromagnetic field. Indeed, potentials of
AB field do not vanish out of the solenoid. AB scattering for spinning
particles was considered in \cite{Hagen91} and \cite{CouPe93} using exact
solutions of Dirac equation in the AB field. A number of theoretical works
and convinced experiments were done to clarify AB effect and to prove its
existence (see, for example, \cite{OlaPo85,Skarz86,PesTo89,Nambu98}).

A progress in study of AB effect may be related to revealing new situations,
where the effect takes place. For example, one can consider more complicated
configurations of electromagnetic fields, different regimes of particle
motions, different dimensions, and so on. To study these new possibilities
one has to have exact solutions of the corresponding quantum equations in
these configurations of electromagnetic fields. In this relation, we ought
to mention exact solutions of the Schr\"{o}dinger equation in a
superposition of AB field and a uniform magnetic field \cite{Lewis83}. The
latter solutions where analyzed in \cite{Sereb85,BagGiS86,SerSk89} from AB
effect point of view. The corresponding coherent states were constructed in 
\cite{BagGiS86}. Klein-Gordon and Dirac equations for particles moving in a
superposition of AB field, Coulomb field, and magnetic monopole field were
found and analyzed in \cite{Villa95,HoaXuK92}.

In this article we present new exact solutions (in 3+1 and 2+1 dimensions)\
of relativistic wave equations (Klein-Gordon and Dirac) in external
electromagnetic fields of special form. This fields are combinations of AB
field and different types of electric and magnetic fields. In Sect.II we
consider AB field combined with longitudinal electromagnetic fields. In
Sect.III superpositions of AB, longitudinal, and crossed fields are studied.
In Sect. IV,V we present solutions in AB field combined with some
non-uniform fields. Here we also discuss some relevant solutions in $2+1$
dimensional QED. Special functions and their properties, which are used in
the article, are present in the Appendix.

Most of works, in which AB effect was studied, are based on the use of exact
solutions of Schr\"{o}dinger equation in AB field \cite{AhaBo59}. Consider
the latter a field in $3+1$ dimensions. If the magnetic solenoid is placed
along the axis $z=x^{3},$ then AB field can be given by potentials (we
denote these potentials as $A_{\mu }^{\left( 0\right) }(x),$ $x=(x^{\mu
},\mu =0,1,2,3)$) of the form 
\begin{equation}
A_{1}^{\left( 0\right) }={\frac{{\Phi }}{{2\pi r^{2}}}}x^{2},\;A_{2}^{\left(
0\right) }=-{\frac{{\Phi }}{{2\pi r^{2}}}}x^{1},\;A_{0}^{\left( 0\right)
}=A_{3}^{\left( 0\right) }=0,\;r^{2}=\left( x^{1}\right) ^{2}+(x^{2})^{2}.
\label{2.1}
\end{equation}
AB magnetic field has the form ${\bf H}^{\left( 0\right) }=(0,0,H^{\left(
0\right) }),$ where $H^{\left( 0\right) }$ is singular at $r=0,$ 
\begin{equation}
H^{\left( 0\right) }=\Phi \delta (x^{1})\delta (x^{2})\,.  \label{2.2}
\end{equation}
AB field creates a finite magnetic flux $\Phi $ along the axis $z$. It is
convenient to define a quantity $\mu ,$ which characterizes the magnetic
flux $\Phi $ and is related to the latter as follows 
\begin{equation}
\Phi =(l_{0}+\mu )\Phi _{0},\;\Phi _{0}=2\pi c\hbar /\left| e\right|
\,,\;0\leq \mu <1,  \label{2.3}
\end{equation}
where $l_{0}$ is integer, and $e=-\left| e\right| $ is the charge of the
electron. In what follows, we call $\mu $ the mantissa of the magnetic flux $%
\Phi .\;$By definition $\mu $ is a positive fractional part of the magnetic
flux if the latter is measured in units of quanta $\Phi _{0}.$\ Cylindrical
coordinates $r,\varphi $ ($x^{1}=r\cos \varphi ,$ $x^{2}=r\sin \varphi $)
are preferable for AB field consideration. In these coordinates 
\begin{equation}
\frac{|e|}{c\hbar }A_{1}^{\left( 0\right) }=\frac{l_{0}+\mu }{r}\sin \varphi
\,,\;\frac{|e|}{c\hbar }A_{2}^{\left( 0\right) }=-\frac{l_{0}+\mu }{r}\cos
\varphi \,\,.  \label{2.8}
\end{equation}

In the present article, we are going to consider particle motion in
electromagnetic fields $A_{\mu }$ that are a combination of AB field and
some additional fields with potentials $A_{\mu }^{\left( 1\right) }$%
\thinspace , 
\begin{equation}
A_{\mu }=A_{\mu }^{\left( 0\right) }+A_{\mu }^{\left( 1\right) }.
\label{2.4}
\end{equation}
Electromagnetic potentials enter in relativistic wave equations only via the
operators of momenta $P_{\mu }=i\hbar \partial _{\mu }-\frac{e}{c}A_{\mu
}\,. $ Doing the transformation $\Psi (x)=e^{-il_{0}\varphi }\tilde{\Psi}(x)$
of wave functions, we can eliminate $l_{0}$ dependence of AB potentials in
equations for $\tilde{\Psi}(x).$ Indeed, such equations already contain
momentum operators of the form 
\begin{eqnarray}
&&e^{il_{0}\varphi }P_{\mu }e^{-il_{0}\varphi }=i\hbar \partial _{\mu }-%
\frac{e}{c}(\tilde{A}_{\mu }^{\left( 0\right) }+A_{\mu }^{\left( 1\right)
}),\;  \nonumber \\
&&\frac{\left| e\right| }{ch}\tilde{A}_{1}^{\left( 0\right) }=\frac{\mu }{r}%
\sin \varphi ,\;\frac{\left| e\right| }{ch}\tilde{A}_{2}^{\left( 0\right) }=-%
\frac{\mu }{r}\cos \varphi ,\;\tilde{A}_{0}^{\left( 0\right) }=\tilde{A}%
_{3}^{\left( 0\right) }=0\,.  \label{2.10}
\end{eqnarray}
Thus, all the matrix elements of any axial-symmetric operators depend on the
mantissa of the magnetic flux only.

\section{Aharonov-Bohm field combined with longitudinal electromagnetic
fields.}

Here we consider particle motion in a superposition of AB field and of some
longitudinal electromagnetic fields. We call electric ${\bf E}$ and magnetic 
${\bf H}$ fields longitudinal ones whenever they are parallel and are
directed along the AB solenoid (along the axis $z$). 
\begin{equation}
{\bf E}=E{\bf n,\;H=}H{\bf n,\;n}^{2}=1\,,\;{\bf n}=(0,0,1)\,.  \label{3.1}
\end{equation}
It follows from Maxwell equations that in this case the functions $E$ and $H$
must obey the conditions 
\begin{eqnarray}
E &=&E\left( x^{0},x^{3}\right) =\partial _{0}A_{3}-\partial
_{3}A_{0},\;A_{0}=A_{0}\left( x^{0},x^{3}\right) ,\;A_{3}=A_{3}\left(
x^{0},x^{3}\right) ;  \nonumber \\
H &=&H\left( x^{1},x^{2}\right) =\partial _{2}A_{1}-\partial
_{1}A_{2},\;A_{1}=A_{1}\left( x^{1},x^{2}\right) ,\;A_{2}=A_{2}\left(
x^{1},x^{2}\right) \,,  \label{3.2}
\end{eqnarray}
where $A_{0}\left( x^{0},x^{3}\right) ,\,A_{1}\left( x^{1},x^{2}\right)
,\,A_{2}\left( x^{1},x^{2}\right) ,\,A_{3}\left( x^{0},x^{3}\right) \;$are
arbitrary functions of the indicated arguments. Exact solutions of the
relativistic wave equations in such fields (in the absence of AB field) were
studied in \cite{BagGiS82,BagGi90}. As we show below, whenever AB field is
present, then exact solutions of the relativistic wave equations can be
found only in the axially symmetric case with the magnetic field having the
form $H=H\left( r\right) .$ Thus, potentials of additional fields, which are
considered in the present Section, have the form: $A_{0}^{\left( 1\right)
}=A_{0}^{\left( 1\right) }\left( x^{0},x^{3}\right) ,\;A_{3}^{\left(
1\right) }=A_{3}^{\left( 1\right) }\left( x^{0},x^{3}\right) $ arbitrary and 
\begin{equation}
\;\;\;A_{1}^{\left( 1\right) }=\frac{c\hbar }{\left| e\right| }\frac{A\left(
r\right) }{r}\sin \varphi ,\;A_{2}^{\left( 1\right) }=-\frac{c\hbar }{\left|
e\right| }\frac{A\left( r\right) }{r}\cos \varphi ,\;H\left( r\right) =\frac{%
c\hbar }{\left| e\right| }\frac{A^{\prime }\left( r\right) }{r}\,,
\label{3.3}
\end{equation}
where $A\left( r\right) $ is an arbitrary function of $r$.

\subsection{Classical description of radial motion.}

To interpret quantum numbers of wave functions, it is often useful to have
classical picture of the problem. That is why we present here a classical
analysis of the particle motion in fields under consideration.

Consider classical trajectories that do not intersect the axis $z$ , thus
they do not ''feel'' the existence of AB field. For such trajectories, the
quantity $P_{r}^{2}$ is an integral of motion ($c^{2}P_{r}^{2}$ is said to
be radial energy), 
\begin{equation}
P_{r}^{2}=P_{1}^{2}+P_{2}^{2}=\hbar
^{2}k_{1}^{2}\,,\;m_{0}^{2}c^{2}+P_{r}^{2}=P_{0}^{2}-P_{3}^{2}\,,
\label{3.1.1}
\end{equation}
where $P_{\mu }$ is the classical kinetic momentum (a classical analog of
the operators $P_{\mu }$) and $m_{0}$\ is the rest mass. $L_{z}$ is an
integral of motion as well ($L$ is the angular momentum), 
\begin{equation}
L_{z}=\tilde{L}_{z}-\hbar \left( l_{0}+\mu \right) =\hbar \left(
l-l_{0}\right) ,\;\tilde{L}_{z}=x^{1}P^{2}-x^{2}P^{1}-\hbar A\left( r\right)
=\hbar \left( l+\mu \right) \,.  \label{3.1.2}
\end{equation}
Here $l$ is arbitrary ($l$ will be integer in quantum theory).

As will be seen below, exact solutions of relativistic wave equations can be
found whenever the functions $A\left( r\right) $in (\ref{3.3}) have the form
\ 
\begin{eqnarray}
1.\;A\left( r\right) &=&0\,,  \label{3.1.5} \\
2.\;A\left( r\right) &=&\frac{\gamma r^{2}}{2}\,,\;\gamma >0\,,
\label{3.1.6} \\
3.\;A\left( r\right) &=&\gamma r\,,\;\gamma >0\,\,.  \label{3.1.7}
\end{eqnarray}
The first case corresponds to the absence of an additional electromagnetic
field, the second one corresponds to the additional constant uniform
magnetic field $H$ along the solenoid $\left( \gamma =\frac{\left| eH\right| 
}{c\hbar }\right) $, and the third one corresponds to the additional
constant magnetic field $H\left( r\right) =b/r,\;\left( \gamma =\frac{\left|
eb\right| }{c\hbar }\right) .\;$Consider classical motion in these cases.

1. For$\;A\left( r\right) =0,$ the momenta $P_{1}$ and $P_{2}$ are integrals
of motion. Then the radial motion (the motion in $x^{1},x^{2}$ plane) is\
parametrized by the proper time $\tau $ and can be presented as 
\begin{equation}
x^{1}=\frac{P^{1}}{m_{0}c}\tau +x_{\left( 0\right) }^{1},\;x^{2}=\frac{P^{2}%
}{m_{0}c}\tau +x_{\left( 0\right) }^{2}\,,  \label{3.1.8}
\end{equation}
where $x_{\left( 0\right) }^{1},x_{\left( 0\right) }^{2}$ are integration
constants. In this case 
\begin{equation}
\tilde{L}_{z}=\hbar \left( l+\mu \right) =x_{\left( 0\right)
}^{1}P^{2}-x_{\left( 0\right) }^{2}P^{1}\,.  \label{3.1.9}
\end{equation}
Consider the quantity 
\begin{equation}
\Delta R=P_{r}^{-1}\left( x_{\left( 0\right) }^{1}P^{2}-x_{\left( 0\right)
}^{2}P^{1}\right) =\frac{l+\mu }{k_{1}}\,.  \label{3.1.10}
\end{equation}
One can show that $\left| \Delta R\right| $ characterizes a minimal distance
between the trajectory (\ref{3.1.8}) and the axis $x^{3}.$ All the classical
trajectories are divided in two groups according to the sign of $l+\mu \,.\;$%
Trajectories with $l+\mu >0$ can be called right ones and those with $l+\mu
<0$ can be called left ones. The reason is the following: Looking from the
positive $z-$direction, one can see that a minimal angle rotation from the
vector ${\bf r}=(x^{1},x^{2},x^{3})$ to the particle momentum is
anticlockwise for the right trajectories and clockwise for left ones.

2. For $A\left( r\right) =\gamma r^{2}/2,$ the radial motion has the form 
\begin{eqnarray}
x^{1} &=&R\cos \kappa +x_{\left( 0\right) }^{1},\;x^{2}=R\sin \kappa
+x_{\left( 0\right) }^{2}\,,  \nonumber \\
\kappa &=&\omega _{0}\tau +\varphi _{0},\;\omega _{0}=\frac{\gamma }{m}%
\,,\;m=\frac{m_{0}c}{\hbar }\,.  \label{3.1.11}
\end{eqnarray}
Here$\ \ R,\varphi _{0},x_{\left( 0\right) }^{1},x_{\left( 0\right) }^{2}$
are integration constants. The trajectories (\ref{3.1.11}) are circles of
radius $R$ with centers having coordinates $x_{\left( 0\right)
}^{1},x_{\left( 0\right) }^{2}\,,$%
\begin{equation}
\left( x^{1}-x_{\left( 0\right) }^{1}\right) ^{2}+\left( x^{2}-x_{\left(
0\right) }^{2}\right) ^{2}=R^{2}\,.  \label{3.1.12}
\end{equation}
One can easily find 
\begin{eqnarray}
&&P_{r}=\hbar \gamma R\,,\;l+\mu \,=\frac{\gamma }{2}\left(
R^{2}-R_{0}^{2}\right) \,,\;\left( x_{\left( 0\right) }^{1}\right)
^{2}+\left( x_{\left( 0\right) }^{2}\right) ^{2}=R_{0\,}^{2},  \nonumber \\
&&\,l+\mu \,\leq \frac{\gamma R^{2}}{2}=\frac{P_{r}^{2}}{2\hbar ^{2}\gamma }=%
\frac{k_{1}^{2}}{2\gamma }\,.  \label{3.1.13}
\end{eqnarray}
We can see that classical trajectories with $l\geq -\mu $ embrace the
solenoid, and ones with $l<-\mu $ do not. In quantum theory these conditions
are $l\geq 0$ and $l<0$ respectively. The quantity $\Delta R$ characterizes
a minimal distance between the trajectory (\ref{3.1.11}) and the solenoid$,$ 
\begin{equation}
\Delta R=\left| R-R_{0}\right| =\frac{2|l+\mu |}{\gamma \left(
R+R_{0}\right) }\,\,.  \label{3.1.14}
\end{equation}
\ 

3. Consider finally $A\left( r\right) =\gamma r\,.$ Here the radial motion
depends essentially on values of constants $a,\varepsilon ,$%
\begin{equation}
a=\frac{P_{r}}{\hbar \gamma }=\frac{k_{1}}{\gamma }>0\,,\;\varepsilon ={\rm %
sign}\left( l+\mu \right) \,.  \label{3.1.15}
\end{equation}
For $\varepsilon =1,$ the classical motion is possible only if $a>1.$ For $%
\varepsilon =-1,$ the classical motion is possible if $a>0$. Whenever $a\geq
1$ we get unbounded motion for $r$. For $0<a<1,$ $\varepsilon =-1,$this
motion is bounded. Below we present the radial motion in $s$ parametrization 
\begin{eqnarray}
&&a>1:\;r=\frac{\left| l+\mu \right| \left( a\cosh s+\varepsilon \right) }{%
\gamma \left( a^{2}-1\right) }\,,\;\tau =\frac{\left| l+\mu \right| \left(
a\sinh s+\varepsilon s\right) }{\gamma ^{2}\left( a^{2}-1\right) ^{3/2}}\,, 
\nonumber \\
&&\varphi -\varphi _{0}=\frac{s}{\sqrt{a^{2}-1}}+2\varepsilon \arctan \left( 
\sqrt{\frac{a-\varepsilon }{a+\varepsilon }}\tanh \frac{s}{2}\right) \,; 
\nonumber \\
&&a=1,\,\varepsilon =-1:\;2\gamma r=\left| l+\mu \right| \left(
s^{2}+1\right) \,,\;2\gamma ^{2}\tau =\left| l+\mu \right| m\left( \frac{%
s^{3}}{3}+s\right) \,,  \nonumber \\
&&\varphi -\varphi _{0}=s-2\arctan s\,;  \nonumber \\
&&a<1,\,\varepsilon =-1:\;r=\frac{\left| l+\mu \right| \left( 1-a\cos
s\right) }{\gamma \left( 1-a^{2}\right) }\,,\;\;\tau =\frac{\left| l+\mu
\right| m\left( s-a\sin s\right) }{\gamma ^{2}\left( 1-a^{2}\right) ^{3/2}}%
\,,  \nonumber \\
&&\varphi -\varphi _{0}=s\left( \frac{1}{\sqrt{1-a^{2}}}-1\right) -2\arctan
\left( \frac{a\sin s}{1+\sqrt{1-a^{2}}-a\cos s}\right) \,.  \label{3.1..16}
\end{eqnarray}

In all the cases under consideration, the minimal distance between a
trajectory and the solenoid is defined by the expression 
\begin{equation}
\Delta R=\frac{\left| l+\mu \right| }{\gamma \left| a-\varepsilon \right| }%
\,.  \label{3.1.17}
\end{equation}
Thus, the quantity $l$ has a clear classical interpretation.

\subsection{Klein-Gordon equation in longitudinal fields}

Here we consider solutions of Klein-Gordon equation 
\begin{equation}
\left( P_{\mu }P^{\mu }-m_{0}^{2}c^{2}\right) \Psi \left( x\right) =0
\label{3.2.1}
\end{equation}
in the superposition of the external fields (\ref{2.1}) and (\ref{3.3}). In
this case, the operators (\ref{3.1.1}) are integrals of motion. Whenever an
additional field is axial-symmetric one (\ref{3.3}), then the operator (\ref
{3.1.2}) is an integral of motion as well. Thus, we subject solutions of the
equation (\ref{3.2.1}) to the following additional conditions 
\begin{equation}
\hbar ^{-2}\left( P_{1}^{2}+P_{2}^{2}\right) \Psi \left( x\right)
=k_{1}^{2}\Psi \left( x\right) \,,\;\hbar ^{-2}\left(
P_{0}^{2}-P_{3}^{2}\right) \Psi \left( x\right) =(m^{2}+k_{1}^{2})\Psi
\left( x\right) \,.  \label{3.2.3}
\end{equation}
Then such solutions can be presented in the form $\Psi \left( x\right) =\psi
\left( x^{1},x^{2}\right) \Phi \left( x^{0},x^{3}\right) \,,$where the
functions $\psi $ and $\Phi $ obey the equations 
\begin{eqnarray}
\hbar ^{-2}\left( P_{1}^{2}+P_{2}^{2}\right) \psi \left( x^{1},x^{2}\right)
&=&k_{1}^{2}\psi \left( x^{1},x^{2}\right) \,,  \label{3.2.5} \\
\;\hbar ^{-2}\left( P_{0}^{2}-P_{3}^{2}\right) \Phi \left(
x^{0},x^{3}\right) &=&(m^{2}+k_{1}^{2})\Phi \left( x^{0},x^{3}\right) \,.
\label{3.2.6}
\end{eqnarray}
AB field does not enter in the equation (\ref{3.2.6}). This equation can be
solved for a large class of electromagnetic fields. All the corresponding
solutions of the equation (\ref{3.2.6})\ are described in detail in \cite
{BagGiS82,BagGi90}, that is why we do not present them here.

Let us turn to the equation (\ref{3.2.5}). As was already mentioned above,
exact solutions of this equation can be found only in the superposition of
AB field and the fields (\ref{3.1.5})-(\ref{3.1.7}). In all these cases, the
operator $L_{z}$ is an integral of motion, thus we can search for solutions
which are eigenvectors for the latter operator. In cylindrical coordinates,
we get 
\begin{equation}
L_{z}\psi \left( r,\varphi \right) =-i\hbar \partial _{\varphi }\psi \left(
r,\varphi \right) =\hbar \left( l-l_{0}\right) \psi \left( r,\varphi \right)
,\;\psi \left( r,\varphi \right) =\frac{\exp \left[ i\left( l-l_{0}\right)
\varphi \right] }{\sqrt{2\pi }}\psi \left( r\right) \,.  \label{3.2.7}
\end{equation}
Whenever additional fields have the structure (\ref{3.3}), the radial
function $\psi \left( r\right) $ obeys the equation 
\begin{equation}
\hat{R}\psi \left( r\right) =k_{1}^{2}\psi \left( r\right) \,,\;\hat{R}%
=\left( \frac{l+\mu +A\left( r\right) }{r}\right) ^{2}-\frac{1}{r}\frac{d}{dr%
}-\frac{d^{2}}{dr^{2}}\,.\;  \label{3.2.8}
\end{equation}

\subsection{Solutions of radial equation in the absence of additional fields}

Consider the equation (\ref{3.2.8}) for $A\left( r\right) =0,$%
\begin{equation}
\frac{d^{2}\psi }{dr^{2}}+\frac{1}{r}\frac{d\psi }{dr}+\left[ k_{1}^{2}-%
\frac{\left( l+\mu \right) ^{2}}{r^{2}}\right] \psi =0\,.\,  \label{3.3.1}
\end{equation}
A general solution of the equation can be written via Bessel functions $%
J_{\nu }\left( x\right) ,$ ( \cite{GraRy94}, 8.402), 
\begin{equation}
\psi \left( r\right) =\psi _{k_{1},l}\left( r\right) =c_{1}J_{\nu }\left(
k_{1}r\right) +c_{2}J_{-\nu }\left( k_{1}r\right) \,,\;\nu =\left| l+\mu
\right| \,.  \label{3.3.2}
\end{equation}
Solutions, which are bounded for all $r\geq 0,$ must have $c_{2}=0.$ In this
case $\psi _{k_{1},l}$ form an orthogonal and complete set of functions ($%
k_{1},k_{1}^{\prime }>0)$, 
\begin{equation}
\int_{0}^{\infty }J_{\nu }\left( k_{1}^{\prime }r\right) J_{\nu }\left(
k_{1}r\right) rdr=k_{1}^{-1}\delta \left( k_{1}-k_{1}^{\prime }\right)
\,,\;\,\int_{0}^{\infty }J_{\nu }\left( k_{1}r^{\prime }\right) J_{\nu
}\left( k_{1}r\right) k_{1}dk_{1}=\frac{\delta \left( r-r^{\prime }\right) }{%
r}.  \label{3.3.3}
\end{equation}
$\,\ \ \ $Suppose $l=0,-1,$ $\mu \neq 0$ (note that in this case the
corresponding classical trajectories pass maximally close to the solenoid).
Then we have a special case. Here there are solutions of the form (\ref
{3.3.2}) with $c_{1}=0,\;c_{2}\neq 0$ (of course they are unbounded), which
obey the relations (\ref{3.3.3}). Moreover, there are solutions (\ref{3.3.2}%
)\ with both $c_{1}\neq 0\;$and $c_{2}\neq 0,$ which are unbounded and not
orthogonal, 
\begin{equation}
\int_{0}^{\infty }J_{\nu }\left( k_{1}^{\prime }r\right) J_{-\nu }\left(
k_{1}r\right) rdr=\frac{2\sin \nu \pi }{\pi \left( k_{1}^{\prime
2}-k_{1}^{2}\right) }\left( \frac{k_{1}^{\prime }}{k_{1}}\right) ^{\nu
}\,,\;k_{1}>k_{1}^{\prime }\,,\;\left| \nu \right| <1\,.  \label{3.3.5}
\end{equation}
However, these solutions are quadratically integrable (due to (\ref{3.3.3}%
))\ and form a complete set of functions (in fact, an overcomplete set).
Besides, the equation (\ref{3.3.1}) has solutions of the form 
\begin{equation}
\psi \left( r\right) =K_{\nu }(qr)\,,\;k_{1}^{2}=-q^{2}\,,  \label{3.3.6}
\end{equation}
where $K_{\nu }(r)$ are Macdonald functions (\cite{GraRy94}, 8.407). These
functions have a finite norm ($\delta =\arg q)$ 
\begin{equation}
\int_{0}^{\infty }K_{\nu }^{\ast }\left( qr\right) K_{\nu }\left( qr\right)
rdr=\frac{\pi \sin 2\nu \delta }{2\left| q\right| ^{2}\sin \nu \pi \sin
2\delta }\,,\;\left| \nu \right| <1\,,\;-\pi /2<\delta <\pi /2\,\;(%
\mathop{\rm Re}%
q>0).  \label{3.3.7}
\end{equation}
But they are not orthogonal with respect to $q,$%
\begin{equation}
\int_{0}^{\infty }K_{\nu }\left( q^{\prime }r\right) K_{\nu }\left(
qr\right) rdr=\frac{\pi \left( q^{2\nu }-q^{\prime 2\nu }\right) }{%
2(q^{2}-q^{\prime 2})\sin \nu \pi }\,,\;\left| \nu \right| <1\,.
\label{3.3.9}
\end{equation}
In particular, for $\nu =0,$ we get 
\begin{equation}
\int_{0}^{\infty }K_{0}^{\ast }\left( qr\right) K_{0}\left( qr\right) rdr=%
\frac{\delta }{\left| q\right| ^{2}\sin 2\delta }\,,\;\int_{0}^{\infty
}K_{0}\left( q^{\prime }r\right) K_{0}\left( qr\right) rdr=\frac{\ln q-\ln
q^{\prime }}{q^{2}-q^{\prime 2}}\,.  \label{3.3.11}
\end{equation}

The above peculiarities are related to the loss of hermicity of the operator 
$\hat{R}$ for $l=0,-1,$ $\mu \neq 0$. Similar problem was discussed in \cite
{Tyuti74}.

\subsection{Uniform magnetic field}

Here we consider a superposition of AB field (\ref{2.1}) and a uniform
magnetic field (\ref{3.1.6}). It is useful to introduce dimensionless
operators $a_{k},a_{k}^{+}$, $k=1,2$ by the relations, 
\begin{eqnarray}
\hbar \sqrt{2\gamma }a_{1} &=&-iP_{1}-P_{2},\;\hbar \sqrt{2\gamma }%
a_{1}^{+}=iP_{1}-P_{2},\;\gamma =\frac{\left| eH\right| }{c\hbar }\,, 
\nonumber \\
\hbar \sqrt{2\gamma }a_{2} &=&-iP_{1}+P_{2}+\hbar \gamma
(x^{1}+ix^{2}),\;\hbar \sqrt{2\gamma }a_{2}^{+}=iP_{1}+P_{2}+\hbar \gamma
(x^{1}-ix^{2})\,.  \label{3.4.1}
\end{eqnarray}
Considering coordinates and momenta in these relations as classical
quantities, we can get a representation for the classical motion (\ref
{3.1.11})\ in terms of $a_{1}$ and $a_{2}$ 
\begin{equation}
a_{1}=\sqrt{\frac{\gamma }{2}}Re^{-i\kappa },\;a_{2}=\sqrt{\frac{\gamma }{2}}%
(x_{0}^{1}+ix_{0}^{2})\,,\;(x_{0}^{1}+ix_{0}^{2})=R_{0}e^{i\delta }\,.
\label{3.4.2}
\end{equation}
The following operator relations take place 
\begin{equation}
P_{r}^{2}=P_{1}^{2}+P_{2}^{2}=\hbar ^{2}\gamma
(a_{1}^{+}a_{1}+a_{1}a_{1}^{+})\,,\;2L_{z}=\hbar
(a_{1}^{+}a_{1}+a_{1}a_{1}^{+}-a_{2}^{+}a_{2}-a_{2}a_{2}^{+}).  \label{3.4.3}
\end{equation}
We introduce also a dimensionless coordinate $\rho $ instead of $r,$ 
\begin{equation}
\rho =\frac{\gamma r^{2}}{2}\,,\;dx^{1}dx^{2}=\frac{1}{\gamma }d\rho
d\varphi \,.  \label{3.4.4}
\end{equation}
On the classical trajectories (\ref{3.1.11}) $\rho $ evolves as 
\begin{equation}
2\rho =\gamma \left[ R^{2}+R_{0}^{2}+2RR_{0}\cos \left( \kappa -\delta
\right) \right] \,.  \label{3.4.5}
\end{equation}
Being written in terms of the variables $\rho ,\varphi ,$ the operators $%
a_{k},a_{k}^{+}$ take the form 
\begin{eqnarray}
&&a_{1}=\sqrt{\rho }e^{-i\varphi }\left[ (l_{0}+\mu +\rho -i\partial
_{\varphi })/2\rho +\partial _{\rho }\right] \;,  \nonumber \\
&&a_{1}^{+}=\sqrt{\rho }e^{i\varphi }\left[ (l_{0}+\mu +\rho -i\partial
_{\varphi })/2\rho -\partial _{\rho }\right] \;,  \nonumber \\
&&a_{2}=-\sqrt{\rho }e^{i\varphi }\left[ (l_{0}+\mu -\rho -i\partial
_{\varphi })/2\rho -\partial _{\rho }\right] \;,  \nonumber \\
&&a_{2}^{+}=-\sqrt{\rho }e^{-i\varphi }\left[ (l_{0}+\mu -\rho -i\partial
_{\varphi })/2\rho +\partial _{\rho }\right] \;.  \label{3.4.6}
\end{eqnarray}
Using the commutation relations for the momentum operators 
\begin{equation}
P_{\mu }P_{\nu }-P_{\nu }P_{\mu }=-i{\frac{e\hbar }{c}}F_{\mu \nu },\;F_{\mu
\nu }=\partial _{\mu }A_{\nu }-\partial _{\nu }A_{\mu }\,,  \label{3.4.7}
\end{equation}
and definition of the magnetic field (\ref{2.2}), we arrive to the following
commutation relations for the operators $a_{k},a_{k}^{+}$ 
\begin{eqnarray}
&[a_{1},a_{1}^{+}]=&1+f,\quad \lbrack a_{2},a_{2}^{+}]=1-f,\quad \lbrack
a_{1},a_{2}]=-f,\quad \lbrack a_{1},a_{2}^{+}]=0\;,  \nonumber \\
&&f=(\Phi /H)\delta (x^{1})\delta (x^{2})=2{\frac{\Phi }{\Phi _{0}}}\delta
(\rho )=2\left( l_{0}+\mu \right) \delta (\rho )\;.  \label{3.4.8}
\end{eqnarray}
These commutation relations contain a singular dimensionless function $f.$
Whenever AB field is absent $(f=0),$ then the operators $a_{k},a_{k}^{+}$
form two mutual commuting sets of creation and annihilation operators. It is
not true in the presence of AB field. However, as it will be seen further,
these operators behave as creation and annihilation ones when acting on
functions that tend to zero (sufficiently rapidly) as $\rho \rightarrow 0$.
Being written in the coordinates $\rho ,\varphi ,$ the operators (\ref{3.4.3}%
) have the form 
\begin{equation}
P_{r}^{2}=2\gamma \hbar ^{2}Q\,,\;L_{z}=-i\hbar \partial _{\varphi \,},\;Q=%
\frac{\left( l_{0}+\mu +\rho -i\partial _{\varphi }\right) ^{2}}{4\rho }%
-\partial _{\rho }-\rho \partial _{\rho }^{2}\,.  \label{3.4.9}
\end{equation}

In the case under consideration, the radial equation (\ref{3.2.8}) reads 
\begin{equation}
\bar{Q}\psi \left( \rho \right) =\left( \bar{n}+\frac{1}{2}\right) \psi
\left( \rho \right) \,,\;\bar{Q}=\frac{\left( l+\mu +\rho \right) ^{2}}{%
4\rho }-\partial _{\rho }-\rho \partial _{\rho }^{2}\,,\;k_{1}^{2}=2\gamma
\left( \bar{n}+\frac{1}{2}\right) \,.  \label{3.4.10}
\end{equation}
Bounded and quadratically integrable solutions of this equation are
expressed via the Laguerre functions (\ref{4.1}) (see Appendix). There are
two types of solutions the latter equation, we denote them as $\psi
^{(j)}(\rho ),\,j=1,2$ (two types of states). The first one $j=1$
corresponds to $l\geq 0$ (classical trajectories embrace the solenoid) 
\begin{equation}
\psi ^{(1)}(\rho )=I_{n+\mu ,n-l}(\rho )\,,\;0\leq l\leq n,\;n=0,1,2,...,\;%
\bar{n}=n+\mu \,.  \label{3.4.11}
\end{equation}
The second type of solutions with $j=2$ corresponds to $l<0$ (classical
trajectories do not embrace the solenoid) 
\begin{equation}
\psi ^{(2)}(\rho )=I_{n-l-\mu ,n}(\rho )\,,\;l<0,\;\bar{n}=n\,.
\label{3.4.12}
\end{equation}
In these two cases radial momentum spectra are different, 
\begin{eqnarray}
\left( k_{1}^{\left( 1\right) }\right) ^{2} &=&2\gamma \left( n+\mu +\frac{1%
}{2}\right) ,\,0\leq l\leq n,  \nonumber \\
\left( k_{1}^{\left( 2\right) }\right) ^{2} &=&2\gamma \left( n+\frac{1}{2}%
\right) ,\;l<0,\;n=0,1,2,...\,.  \label{3.4.13}
\end{eqnarray}
The spectrum for $j=2$ (which is a part of the total spectrum) corresponds
exactly to the spectrum of a spinless particle in a uniform magnetic field
(without AB\ field). The spectrum for $j=1$ is shifted by $2\gamma \mu $
with respect to the one for $j=2$. It is important to note that the presence
of AB field lifts partially the degeneracy of the total spectrum in the
quantum number $l.$

It is convenient to define effective quantum numbers $\bar{l}\;$and$\;\bar{n}
$ by the relations 
\begin{equation}
\bar{n}=n+\mu \left( 2-j\right) =\left\{ 
\begin{array}{c}
n+\mu \,,\;j=1,\;\bar{l}=l+\mu ,\;\bar{l}\leq \bar{n} \\ 
n\,,\;j=2,\;n=0,1,2,...\,.
\end{array}
\right.  \label{3.4.14}
\end{equation}
Using these numbers, we introduce the functions 
\begin{eqnarray}
&&\psi _{n,l}^{(1)}(\rho ,\varphi )=\left( -1\right) ^{n-l}\frac{\exp \left[
i(l-l_{0})\varphi \right] }{\sqrt{2\pi }}I_{\bar{n},\bar{n}-\bar{l}}(\rho
)\,,  \nonumber \\
&&\psi _{n,l}^{(2)}(\rho ,\varphi )=\left( -1\right) ^{n}\frac{\exp \left[
i(l-l_{0})\varphi \right] }{\sqrt{2\pi }}I_{\bar{n}-\bar{l},\bar{n}}(\rho
)\,.  \label{3.4.15}
\end{eqnarray}
According to (\ref{4.4}), these functions can be expressed via the Laguerre
polynomials. Thus, the orthonormality relation can be proved 
\begin{equation}
\int_{0}^{\infty }d\rho \int_{0}^{2\pi }d\varphi \,\psi _{n^{\prime
},l^{\prime }}^{(j^{\prime })\ast }(\rho ,\varphi )\psi _{n,l}^{(j)}(\rho
,\varphi )=\delta _{l.l^{\prime }}\delta _{n,n^{\prime }}\,.  \label{3.4.16}
\end{equation}
The set of the Laguerre functions 
\begin{equation}
I_{\alpha +n,n}(x),\;n=0,1,2...\,\,,\;\alpha >-1  \label{3.4.17}
\end{equation}
is complete in the space of quadratically integrable functions of $\ x\geq
0, $%
\begin{equation}
\sum_{n=0}^{\infty }I_{\alpha +n,n}(x)I_{\alpha +n,n}(y)=\delta \left(
x-y\right) \,.  \label{3.4.18}
\end{equation}
Then the set $\psi _{n,l}^{(j)}(\rho ,\varphi )$ is complete in the space of
quadratically integrable functions of $\rho ,\varphi ,$ ($\rho >0,$\ $0\leq
\varphi \leq 2\pi $).

Using the relations (\ref{4.11})-(\ref{4.16}), one can get the action of the
operators (\ref{3.4.6}) on the functions $\psi _{n,l}^{(j)}(\rho ,\varphi ),$%
\begin{eqnarray}
a_{1}\psi _{n,l}^{(j)}(\rho ,\varphi ) &=&\sqrt{\bar{n}}\psi
_{n-1,l-1}^{(j)}(\rho ,\varphi ),\;a_{1}^{+}\psi _{n,l}^{(j)}(\rho ,\varphi
)=\sqrt{\bar{n}+1}\psi _{n+1,l+1}^{(j)}(\rho ,\varphi )\,,  \nonumber \\
a_{2}\psi _{n,l}^{(j)}(\rho ,\varphi ) &=&\sqrt{\bar{n}-\bar{l}}\psi
_{n,l+1}^{(j)}(\rho ,\varphi ),\;a_{2}^{+}\psi _{n,l}^{(j)}(\rho ,\varphi )=%
\sqrt{\bar{n}-\bar{l}+1}\psi _{n,l-1}^{(j)}(\rho ,\varphi )\,.
\label{3.4.19}
\end{eqnarray}
These formulas show that the functions $\psi _{n,l}^{(1)}$ may be created by
an action of the operators $a_{k}^{+}$ on $\psi _{0,0}^{(1)}\;,$ and the
functions $\psi _{n,l}^{(2)}$ may be created by an action of the operators $%
a_{k}^{+}$ on $\psi _{0,-1}^{(2)}\;.$ Namely, 
\begin{eqnarray}
&&\psi _{n,l}^{(1)}=\sqrt{{\frac{{\Gamma (1+\mu )}}{{\Gamma (1+\bar{n}%
)\Gamma (1+\bar{n}-\bar{l})}}}}{(a_{2}^{+})}^{n-l}{(a_{1}^{+})}^{n}\psi
_{0,0}^{(1)}\;,  \label{3.4.20} \\
&&\psi _{n,l}^{(2)}=\sqrt{{\frac{{\Gamma (2-\mu )}}{{\Gamma (1+\bar{n}%
)\Gamma (1+\bar{n}-\bar{l})}}}}{(a_{1}^{+})}^{n}{(a_{2}^{+})}^{n-l-1}\psi
_{0,-1}^{(2)}\;.  \label{3.4.21}
\end{eqnarray}
It is natural to interpret $\psi _{0,0}^{(1)}$ as a vacuum state for the
states $\psi _{n,l}^{(1)}\;,$ and to interpret $\psi _{0,-1}^{(2)}$ as a
vacuum state for the states $\psi _{n,l}^{(2)}\;.$ Thus, for $\mu \neq 0,$
we have two vacuum states in the problem. For $\mu =0,$ the situations
changes. By virtue of (\ref{4.25}) 
\begin{equation}
I_{n,n-l}=\left( -1\right) ^{l}I_{n-l,n}\;\rightarrow \psi _{n,l}^{(1)}={(-1)%
}^{l}\psi _{n,l}^{(2)}\,,\;\;\mu =0\;,  \label{3.4.23}
\end{equation}
and for any $l<n$, the function $\psi _{0,0}^{(1)}$ is connected to $\psi
_{0,-1}^{(2)}$ as 
\begin{equation}
a_{2}^{+}\psi _{0,0}^{(1)}=\psi _{0,-1}^{(2)},\quad a_{2}\psi
_{0,-1}^{(2)}=\psi _{0,0}^{(1)}\;.  \label{3.4.22}
\end{equation}
Thus, we have the only one vacuum in the problem, one energy spectrum (\ref
{3.4.13}), and all the wave functions are created from the vacuum $\psi
_{0,0}^{(1)}$.

One ought to stress that all the states obey the property $\psi
_{n,l}^{(j)}\left( \rho =0,\varphi \right) =0\,,$ which means that the
scalar particle has zero probability to be found in the solenoid area. In
fact, the existence of this property allows us to speak about AB effect.

The definitions (\ref{3.4.19}) can formally be considered for any values of
indices $n,$ $l.$ In particular, we can consider the following relations 
\begin{eqnarray}
&&a_{1}^{+}\psi _{n,-1}^{(2)}=\sqrt{n+1}\psi _{n+1,0}^{(2)}  \nonumber \\
&=&\left( -1\right) ^{n+1}(1+n)\sqrt{\frac{{\Gamma (1+n)}}{2\pi {\Gamma
(2-\mu +n)}}}\exp [-il_{0}\varphi -{\frac{\rho }{2}}]\rho ^{-{\frac{\mu }{2}}%
}L_{n+1}^{-\mu }(\rho ),\;  \nonumber \\
&&a_{1}\psi _{n,0}^{(1)}=\sqrt{n+\mu }\psi _{n-1,-1}^{(1)}  \nonumber \\
&=&\left( -1\right) ^{n}(n+\mu )\sqrt{\frac{{\Gamma (1+n)}}{2\pi {\Gamma
(1+\mu +n)}}}\exp [-i(1+l_{0})\varphi -{\frac{\rho }{2}}]\rho ^{-{\frac{{%
1-\mu }}{2}}}L_{n}^{\mu -1}(\rho ),  \nonumber \\
&&a_{2}^{+}\psi _{n,0}^{(1)}=\sqrt{n+1}\psi _{n,-1}^{(1)}  \nonumber \\
&=&\left( -1\right) ^{n+1}(1+n)\sqrt{\frac{{\Gamma (1+n)}}{2\pi {\Gamma
(1+\mu +n)}}}\exp [-i(1+l_{0})\varphi -{\frac{\rho }{2}}]\rho ^{-{\frac{{%
1-\mu }}{2}}}L_{n+1}^{\mu -1}(\rho ),  \nonumber \\
&&a_{2}\psi _{n,-1}^{(2)}=\sqrt{1-\mu +n}\psi _{n,0}^{(2)}  \nonumber \\
&=&\left( -1\right) ^{n}(1-\mu +n)\sqrt{\frac{{\Gamma (1+n)}}{2\pi {\Gamma
(2-\mu +n)}}}\exp [-il_{0}\varphi -{\frac{\rho }{2}}]\rho ^{-{\frac{\mu }{2}}%
}L_{n}^{-\mu }(\rho ).  \label{3.4.25}
\end{eqnarray}
However, the functions $\psi _{n,-1}^{(1)},\;\psi _{n,0}^{(2)}$ do not
present any physical solutions of the problem, they are not in the set (\ref
{3.4.15}). Thus, in the general case $\mu \neq 0$, the action of the
operators $a_{k}^{+},a_{k}$ on wave functions may lead them out of a class
of physical solutions. The functions (\ref{3.4.25}) are singular at $r=0$
(for $\mu \neq 0$), however they still remain quadratically integrable.

Thus, we see that $l=0,-1$ is a special case. Here there appear unbounded
(but quadratically integrable) solution $\psi _{n,-1}^{(1)},\;\psi
_{n,0}^{(2)}.$ Whenever $\mu \rightarrow 0,$ these states either coincide
with the corresponding states in the pure magnetic field or disappear. The
states $\psi _{n,0}^{(1)},\,\psi _{n,0}^{(2)},\,\psi _{n,-1}^{(1)},\;\psi
_{n,-1}^{(2)}$ are not mutually orthogonal in spite of the fact that they
belong to different eigenvalues of the operator $P_{r}^{2}\,.$

The equation (\ref{3.4.10}) has additional solutions in the case $l=0,-1.$
According to (\ref{4.29}) they have the form 
\begin{equation}
\psi \left( \rho \right) =\psi _{\lambda \alpha }\left( \rho \right)
,\;\alpha =\left\{ 
\begin{array}{c}
\mu ,\;l=0 \\ 
1-\mu ,\;l=-1
\end{array}
\right. ,\;\;2\bar{n}=2\lambda +l+\mu -1,  \label{3.4.26}
\end{equation}
where the functions $\psi _{\lambda \alpha }\left( \rho \right) $ are
defined by Eqs. (\ref{4.139}), (\ref{4.140}). Solutions (\ref{3.4.26}) with
any different complex $\lambda $ are orthogonal and have finite norms
according to the properties (\ref{4.148})-(\ref{4.150}). These solutions are
singular at $r=0.$ For $l=0$ such solutions exist even in the pure magnetic
field. Their existence is related to the loss of hermicity of the operator $%
P_{r}^{2}.$

\subsection{Nonuniform magnetic field}

Here we consider the radial equation (3.2.8) for $A\left( r\right) =\gamma r 
$, 
\begin{equation}
\frac{d^{2}\psi }{dr^{2}}+\frac{1}{r}\frac{d\psi }{dr}-\left( \frac{l+\mu
+\gamma r}{r}\right) ^{2}\psi +\left( \gamma a\right) ^{2}\psi
=0,\;P_{r}^{2}=\left( \hbar \gamma a\right) ^{2},\;k_{1}^{2}=\left( \gamma
a\right) ^{2}\,.  \label{3.5.1}
\end{equation}
The constant $a$ is defined in (\ref{3.1.15}).

For $a\neq 1$ bounded solutions of this equation are expressed via the the
Laguerre functions (\ref{4.1}) as 
\begin{eqnarray}
\psi \left( r\right) &=&\psi _{n,l}\left( r\right) =I_{\alpha
+n,n}(x)\,,\;x=2\sqrt{1-a^{2}}\gamma r\,,\;  \nonumber \\
\alpha &=&2\left| l+\mu \right| \,,\;1+\alpha +2n=-\frac{2\left( l+\mu
\right) }{\sqrt{1-a^{2}}}.  \label{3.5.2}
\end{eqnarray}

For $a>1,$ there are solutions\ for any $l$ (in complete accordance with the
classical theory). In this case the Laguerre functions have imaginary
arguments and complex indices.

For $a<1,$ bounded solutions of the form (\ref{3.5.2}) exist only for $l<0$
(also in accordance with the classical theory). Besides, in such a case 
\begin{equation}
a^{2}=1-\frac{\alpha ^{2}}{\left( 1+\alpha +2n\right) ^{2}}%
\,,\;n=0,1,2,...\;;\;x=\frac{2\alpha \gamma r}{1+\alpha +2n}\,.
\label{3.5.3}
\end{equation}
Thus, $n$ must be integer and $a$ is quantized. Here the functions (\ref
{3.5.2}) can be expressed via the Laguerre polynomials by means of (\ref{4.4}%
), (\ref{4.28}).

For $a=1,$ bounded solutions of the form (\ref{3.5.2}) exist only for $l<0$,
these functions can be expressed via the Bessel functions, 
\begin{equation}
\psi \left( r\right) =J_{\alpha }\left( 2\sqrt{\alpha \gamma r}\right)
\,,\;a=1,\;l<0\,.  \label{3.5.5}
\end{equation}
One can see with the help of (\ref{4.40}) that the solutions (\ref{3.5.5})\
follow from (\ref{3.5.2}) as $a\rightarrow 1$.

All the bounded solutions vanish at $r=0.$

It is interesting to note that there exist unbounded (but quadratically
integrable solutions) of the equation (\ref{3.5.1}) for $l=0,-1$ . For any
complex $a,$ the latter solutions are defined as 
\begin{eqnarray}
\psi \left( r\right) &=&\psi _{\lambda ,\alpha }\left( 2\sqrt{1-a^{2}}\gamma
r\right) ,\;a\neq 1,\;\lambda =-\frac{l+\mu }{\sqrt{1-a^{2}}}\,,\;%
\mathop{\rm Re}%
\sqrt{1-a^{2}}>0\,,  \nonumber \\
\psi \left( r\right) &=&K_{\alpha }\left( 2\sqrt{\alpha \gamma r}\right)
\,,\;a=1\,.  \label{3.5.6}
\end{eqnarray}
Here $K_{\alpha }\left( x\right) $ are Macdonald functions. The existence of
such solutions is related to the loss of hermicity of the operator $%
P_{r}^{2}\,.$

\subsection{Solutions of Klein-Gordon equation that are not related to
radial momentum conservation}

As was demonstrated above, selecting the radial momentum (\ref{3.2.5}) as an
integral of motion, we can separate variables and then consider two
independent problems: a two-dimensional motion of the charge in the magnetic
field (\ref{3.2.5}) (the latter field includes AB field), and a
two-dimensional motion of the charge in an electric field, the latter
problem does not depend on AB field. However, there is a wide class of exact
solutions, which are not eigenvectors for the radial momentum operator. They
correspond to a superposition of AB field and longitudinal running electric
fields (potentials of such fields depend on $u^{0}=x^{0}-x^{3}$ only). Thus,
here we will use light cone variables $u^{0},u^{3},$%
\begin{equation}
u^{0}=x^{0}-x^{3}\,,\;u^{3}=x^{0}+x^{3}\,.  \label{3.6.1}
\end{equation}
\ Then the above mentioned longitudinal running electric fields have the
following potentials and strengths 
\begin{equation}
A_{0}^{\left( 1\right) }=A_{3}^{\left( 1\right) }=\frac{1}{2}B\left(
u^{0}\right) ,\;E=B^{\prime }\left( u^{0}\right) \,.  \label{3.6.2}
\end{equation}
where $B\left( u^{0}\right) $ is an arbitrary function of $u^{0}.$ Consider
operators $\tilde{P}_{0}\,,\tilde{P}_{3}\,,\tilde{p}_{3}\,,$%
\begin{eqnarray}
&&2\tilde{P}_{0}\,=P_{0}-P_{3},\;2\tilde{P}_{3}=P_{0}+P_{3},\,\tilde{P}%
_{0}\,=i\hbar \frac{\partial }{\partial u^{0}}\,,  \nonumber \\
&&\tilde{p}_{3}=i\hbar \frac{\partial }{\partial u^{3}}\,,\;\tilde{P}_{3}=%
\tilde{p}_{3}+\frac{\hbar g\left( u^{0}\right) }{2}\,,\;g=\frac{\left|
e\right| B\left( u^{0}\right) }{c\hbar }\,.  \label{3.6.4}
\end{eqnarray}
The operator $\tilde{p}_{3}$ commutes with the one $L_{z}$ and both are
integrals of motion. Thus, we can demand for solutions of Eq. (\ref{3.2.1})
to be eigenvectors for these operators, 
\begin{equation}
\tilde{p}_{3}\Psi \left( x\right) =\frac{\hbar \lambda }{2}\Psi \left(
x\right) \,,\;L_{z}\Psi \left( x\right) =\hbar \left( l-l_{0}\right) \Psi
\left( x\right) \,.  \label{3.6.5}
\end{equation}
Such solutions have the form 
\begin{eqnarray}
\Psi \left( x\right) &=&\left[ \lambda +g\left( u^{0}\right) \right]
^{-1/2}\Phi \left( r,t\right) \exp i\left[ \left( l-l_{0}\right) \varphi
-m^{2}t\left( u^{0}\right) -\frac{\lambda u^{3}}{2}\right] ,  \nonumber \\
t\left( u^{0}\right) &=&\frac{1}{2}\int \frac{du^{0}}{\lambda +g\left(
u^{0}\right) }\,,  \label{3.6.7}
\end{eqnarray}
where the function $\Phi \left( r,t\right) $ obeys the equation 
\begin{equation}
\hat{R}_{1}\Phi \left( r,t\right) =0\,,\;\hat{R}_{1}=i\partial _{t}+\partial
_{r}^{2}+\frac{\partial _{r}}{r}-\frac{\left[ \bar{l}+A\left( r\right) %
\right] ^{2}}{r^{2}}\,\,,\;\,\bar{l}=l+\mu \;\,.  \label{3.6.8}
\end{equation}
We recall that $A\left( r\right) $ was defined in (\ref{3.3}).

Consider first the case $A\left( r\right) =0.$ Here we find a propagation
function for the equation (\ref{3.6.8}) in the form ($J_{\nu }\left(
x\right) $ are the Bessel functions) 
\begin{eqnarray}
&&G_{0}\left( r,r^{\prime },t\right) =\frac{1}{2t}J_{\left| \bar{l}\right|
}\left( \frac{rr^{\prime }}{2t}\right) e^{iQ_{0}}\,,\;Q_{0}=\frac{%
r^{2}+r^{\prime 2}}{4t}-\frac{\left( \left| \bar{l}\right| +1\right) \pi }{2}%
\,,  \nonumber \\
&&\left. \hat{R}_{1}\right| _{A=0}G_{0}\left( r,r^{\prime },t\right)
=0\,,\;\lim_{t\rightarrow 0}G_{0}\left( r,r^{\prime },t\right) =\frac{1}{r}%
\delta \left( r-r^{\prime }\right) \,.  \label{3.6.9}
\end{eqnarray}
The case $A\left( r\right) =\rho =\gamma r^{2}/2$ can be considered in the
same manner. Here the propagation function has the form 
\begin{eqnarray}
&&G\left( \rho ,\rho ^{\prime },t\right) =\frac{1}{2\sin \tau }J_{\left| 
\bar{l}\right| }\left( \frac{\sqrt{\rho \rho ^{\prime }}}{\sin \tau }\right)
e^{iQ}\,,\;Q=\frac{\rho +\rho ^{\prime }}{\sin \tau }-\frac{\left( \left| 
\bar{l}\right| +1\right) \pi }{2}-\bar{l}\tau \,,  \nonumber \\
&&\hat{R}_{1}G\left( \rho ,\rho ^{\prime },t\right)
=0\,,\;\lim_{t\rightarrow 0}G\left( \rho ,\rho ^{\prime },t\right) =\delta
\left( \rho -\rho ^{\prime }\right) \,,\;\tau =\gamma t\left( u^{0}\right)
\,.  \label{3.6.12}
\end{eqnarray}
The functions $G_{0}\left( r,r^{\prime },t\right) $ and $G\left( \rho ,\rho
^{\prime },t\right) $ solve the Cauchy problem. For example, 
\begin{equation}
\Phi \left( \rho ,t\right) =\int_{0}^{\infty }G\left( \rho ,\rho ^{\prime
},t\right) \Phi \left( \rho ^{\prime }\right) d\rho ^{\prime },
\label{3.6.14}
\end{equation}
where $\Phi \left( \rho \right) $ is an arbitrary functions (an initial date
for $\Phi \left( \rho ,t\right) $).

For the field (\ref{3.1.7}), the corresponding propagation function is quite
complicated \cite{KomPoS76}.

\subsection{Exact solutions of Dirac equation}

Here we are going to study Dirac equation 
\begin{equation}
(\gamma ^{\mu }P_{\mu }-m_{0}c)\Psi \left( x\right) =0\,  \label{3.7.1}
\end{equation}
in the superposition of AB field and the field (\ref{3.1}). We use a
standard representation (see for example \cite{BagGi90}) for $\gamma -$%
matrices. In the case under consideration, we look for solutions with a
definite radial momentum. The corresponding bispinors $\Psi \left( x\right) $
can be written in a block form 
\begin{equation}
\Psi \left( x\right) =Q\left( 
\begin{array}{c}
\psi _{1}\left( x^{1},x^{2}\right) \left[ m+F-ik_{1}\sigma _{2}\right]  \\ 
\psi _{2}\left( x^{1},x^{2}\right) \left[ \left( m-F\right) \sigma
_{3}-ik_{1}\sigma _{1}\right] 
\end{array}
\right) \upsilon \tilde{\Phi}\left( x^{0},x^{3}\right) \,,\;F=\hbar
^{-1}\left( P_{0}+P_{3}\right) \,,  \label{3.7.2}
\end{equation}
where $\upsilon $ is an arbitrary spinor; $\sigma _{k}\;(k=1,2,3)$\ are
Pauli matrices; the function $\tilde{\Phi}\left( x^{0},x^{3}\right) $ obeys
the equation 
\begin{equation}
\left[ \hbar ^{-2}\left( P_{0}^{2}-P_{3}^{2}\right) +i\frac{\left| e\right| E%
}{c\hbar }\right] \tilde{\Phi}\left( x^{0},x^{3}\right) =(m^{2}+k_{1}^{2})%
\tilde{\Phi}\left( x^{0},x^{3}\right) ,  \label{3.7.3}
\end{equation}
where $E=E\left( x^{0},x^{3}\right) $ is electric field strength (\ref{3.2}%
), and functions $\psi _{1},\,\psi _{2}$ obey the following equations 
\begin{equation}
\left( P_{1}+iP_{2}\right) \psi _{1}\left( x^{1},x^{2}\right) =\hbar
k_{1}\psi _{2}\,\left( x^{1},x^{2}\right) ,\;\left( P_{1}-iP_{2}\right) \psi
_{2}\left( x^{1},x^{2}\right) =\hbar k_{1}\psi _{1}\left( x^{1},x^{2}\right)
\,.  \label{3.7.7}
\end{equation}
The presence of the arbitrary spinor $\upsilon $ in the solutions (\ref
{3.7.2}) indicates that Eq. (\ref{3.7.1}) does not fix the spin orientation.
This orientation can be fixed by a choice of a spin operator \cite
{BagGiS82,BagGi90}. One has to stress that the equation (\ref{3.7.3}) does
not contain AB field. All possible exact solutions of this equation were
presented in \cite{BagGiS82,BagGi90}, thus here we do not repeat these
results.

The fields (\ref{3.3}) are axially symmetric, thus $J_{z}$ is an integral of
motion in such a case ($J\;$is total angular momentum operator). Let us
consider solutions that are eigenvectors for this operator, 
\begin{equation}
J_{z}\Psi =\hbar \left( l-l_{0}-\frac{1}{2}\right) \Psi \,,\;J_{z}=L_{z}+%
\frac{\hbar }{2}\Sigma _{3}\,,\;l=0,\pm 1,\pm 2,...\,\,.  \label{3.7.8}
\end{equation}
(${\bf \Sigma =}{\rm diag}\left( {\bf \sigma ,\sigma }\right) \,).$We obey
the equations (\ref{3.7.8}) choosing 
\begin{equation}
\psi _{1}\left( x^{1},x^{2}\right) =\frac{\exp \left[ i\left(
l-l_{0}-1\right) \varphi \right] }{\sqrt{2\pi }}\psi _{1}\left( r\right)
\,,\;\psi _{2}\left( x^{1},x^{2}\right) =-i\frac{\exp \left[ i\left(
l-l_{0}\right) \varphi \right] }{\sqrt{2\pi }}\psi _{2}\left( r\right) \,,
\label{3.7.9}
\end{equation}
where the functions $\psi _{k}\left( r\right) $ satisfy a set of first order
differential equations 
\begin{equation}
\left( \frac{\bar{l}+A\left( r\right) }{r}+\frac{d}{dr}\right) \psi
_{2}\left( r\right) =k_{1}\psi _{1}\left( r\right) \,,\;\left( \frac{\bar{l}%
-1+A\left( r\right) }{r}-\frac{d}{dr}\right) \psi _{1}\left( r\right)
=k_{1}\psi _{2}\left( r\right) \,.  \label{3.7.10}
\end{equation}
Consider solutions of the latter equations for the fields (\ref{3.1.5})-(\ref
{3.1.7}).

For $A\left( r\right) =0,$ we deal with the pure AB field. For $l\neq 0,$
all bounded solutions of Eq. (\ref{3.7.10}) have the form 
\begin{equation}
\psi _{1}\left( r\right) =J_{\left| \bar{l}-1\right| }\left( k_{1}r\right)
\,,\;\psi _{2}\left( r\right) =\varepsilon J_{\left| \bar{l}\right| }\left(
k_{1}r\right) \,,\;\varepsilon ={\rm sign\,}l\,,  \label{3.7.11}
\end{equation}
where $J_{\mu }\left( x\right) $ are the Bessel functions. These solutions
vanish at $r=0.$ For $l=0,\;\mu \neq 0,$ the system of equations (\ref
{3.7.10}) has no bounded solutions. In such a case, a general solution of
this system has the form 
\begin{equation}
\psi _{1}\left( r\right) =c_{1}J_{\mu -1}\left( k_{1}r\right) +c_{2}J_{1-\mu
}\left( k_{1}r\right) \,,\;\psi _{2}\left( r\right) =c_{1}J_{\mu }\left(
k_{1}r\right) -c_{2}J_{-\mu }\left( k_{1}r\right) \,.  \label{3.7.12}
\end{equation}
where $c_{1},c_{2}$ are arbitrary constants. In spite of the fact that these
solutions are unbounded they still are quadratically integrable (as in the
scalar case). Moreover, for any complex $k_{1}$ $\left( 
\mathop{\rm Re}%
k_{1}>0\right) $ there exist unbounded solutions with a finite norm, they
are expressed via the Macdonald functions, 
\begin{equation}
\psi _{1}\left( r\right) =K_{1-\mu }\left( k_{1}r\right) \,,\;\psi
_{2}\left( r\right) =-K_{\mu }\left( k_{1}r\right) \,,\;0<\mu <1\,.
\label{3.7.13}
\end{equation}
Similar to the scalar case, we can conclude that\ the operator $\gamma ^{\mu
}P_{\mu }$ is not self-conjugate anymore for $l=0,\,\mu \neq 0$. In contrast
to the scalar case, there are no quadratically integrable unbounded
solutions for $l=-1,$ as well as for $l=0,$ $\mu =0.$

Consider now the case of the uniform magnetic field (\ref{3.1.6}). Using the
operators (\ref{3.4.6}), we can write the equations (\ref{3.7.7}) as 
\[
a_{1}\psi _{2}\left( \rho ,\varphi \right) =-i\sqrt{\bar{n}}\psi _{1}\left(
\rho ,\varphi \right) \,,\;a_{1}^{+}\psi _{1}\left( \rho ,\varphi \right) =i%
\sqrt{\bar{n}}\psi _{2}\left( \rho ,\varphi \right) \,,\;k_{1}=\sqrt{2\gamma 
\bar{n}}\,,\;\rho =\frac{\gamma r^{2}}{2}\,. 
\]
Their solutions have the form (see (\ref{3.4.19}), (\ref{3.4.15})) 
\begin{equation}
\psi _{1}\left( \rho ,\varphi \right) =\psi _{n-1,l-1}^{\left( j\right)
}\left( \rho ,\varphi \right) \,,\;\psi _{2}\left( \rho ,\varphi \right)
=-i\psi _{n,l}^{\left( j\right) }\left( \rho ,\varphi \right) \,.
\label{3.7.15}
\end{equation}
As in the scalar case, there are two types of states (with $j=1,2$). These
states are bounded at $l\neq 0$; they vanish at $r=0.$ The states (\ref
{3.7.15}) are unbounded at $l=0$ but they still are quadratically
integrable. Besides, there are unbounded solutions with finite norms for any
complex $\bar{n}.$ Such solutions are expressed via the functions $\psi
_{\lambda ,\alpha }\left( x\right) $ (the latter are defined by (\ref{4.139}%
), (\ref{4.140})) as 
\begin{equation}
\psi _{1}\left( r\right) =\bar{n}^{3/4}\psi _{\lambda -\frac{1}{2},1-\mu
}\left( \rho \right) \,,\;\psi _{2}\left( r\right) =\bar{n}^{1/4}\psi
_{\lambda ,\mu }\left( \rho \right) \,,\;2\bar{n}=2\lambda +\mu -1\,.
\label{3.7.16}
\end{equation}
Thus, we see that the operator $\gamma ^{\mu }P_{\mu }$ is not
self-conjugate for $l=0,\,\mu \neq 0$ as well. All the above singular
solutions vanish or become nonsingular as $\mu \rightarrow 0.$

One ought also remark that $\psi _{1}$ (which correspond to $j=2)$ from (\ref
{3.7.15}) vanish at $n=0.$ Thus, the complete wave function (\ref{3.7.2}) is
an eigenvector for the operator $\Sigma _{3},$ 
\begin{equation}
\Sigma _{3}\Psi =-\Psi \,.  \label{3.7.17}
\end{equation}
That means that in such states the electron spin has the only one
orientation, namely, opposite to the magnetic field.

Consider finally the case of nonuniform magnetic field (\ref{3.1.7}). For $%
l\neq 0,$\ $a\neq 1,$ the corresponding bounded solutions (they also vanish
at $r=0$) have the form 
\begin{eqnarray}
\psi _{1}\left( r\right) &=&I_{n-1,n+1-2\bar{l}}\left( x\right) \,,\;\psi
_{2}\left( r\right) =-I_{n,n-2\bar{l}}\left( x\right) \,,\;l>0\,;  \nonumber
\\
\,\psi _{1}\left( r\right) &=&I_{n+1-2\bar{l},n-1}\left( x\right) \,,\;\psi
_{2}\left( r\right) =-I_{n-2\bar{l},n}\left( x\right) \,,\;l<0\,;  \nonumber
\\
x &=&2\sqrt{1-a^{2}}\gamma r\,,\;1-2\bar{l}+2n=\frac{1-2\bar{l}}{\sqrt{%
1-a^{2}}}\,,\;\bar{l}=l+\mu \,,  \label{3.7.18}
\end{eqnarray}
where $I_{n,m}\left( x\right) $ are the Laguerre functions (\ref{4.1}), and
we use the notation (\ref{3.1.15}). Whenever $a^{2}>1,$ any $l\neq 0$ are
admissible in complete agreement with classical theory. Whenever $a^{2}<1,$
the only $l<0$ are admissible. In such a case $n$ is integer and the
functions (\ref{3.7.18}) are expressed via the Laguerre polynomials
according to (\ref{4.28}). At the same time, the following quantization
takes place 
\begin{equation}
a^{2}=1-\frac{\left( 1+2\left| \bar{l}\right| \right) ^{2}}{\left( 1+2\left| 
\bar{l}\right| +2n\right) ^{2}}\,,\;n=0,1,2,...\;.  \label{3.7.19}
\end{equation}
For $a=1,$ $l\neq 0,$ the only bounded states can be found for $l<0.$ They
are expressed via the Bessel functions, 
\begin{equation}
\psi _{1}\left( r\right) =J_{2\left| \bar{l}\right| +2}\left( 2\sqrt{\left(
1+2\left| \bar{l}\right| \right) \gamma r}\right) \,,\;\psi _{2}\left(
r\right) =-J_{2\left| \bar{l}\right| }\left( 2\sqrt{\left( 1+2\left| \bar{l}%
\right| \right) \gamma r}\right) \,.  \label{3.7.20}
\end{equation}
Solutions (\ref{3.7.20}) follow from (\ref{3.7.18}) as $a\rightarrow 1.$
That fact can be confirmed by the use of the limit (\ref{4.40}).

$l=0$ is a special case. Here there are only unbounded solutions. Some of
them are quadratically integrable. Whenever $a^{2}>1,$ such solutions have
the form 
\begin{eqnarray}
\psi _{1}\left( r\right) &=&c_{1}I_{n+1-2\mu ,n-1}\left( x\right)
+c_{2}I_{n-1,n+1-2\mu }\left( x\right) \,,  \nonumber \\
\psi _{2}\left( r\right) &=&-c_{1}I_{n-2\mu ,n}\left( x\right)
-c_{2}I_{n,n-2\mu }\left( x\right) \,,  \label{3.7.21}
\end{eqnarray}
where $c_{k}$ are arbitrary constants, and for $a=1$ these solutions read 
\begin{eqnarray}
\psi _{1}\left( r\right) &=&J_{2-2\mu }\left( z\right) ,\;\psi _{2}\left(
r\right) =-J_{-2\mu }\left( z\right) ,\;0<\mu <\frac{1}{2}\,;  \nonumber \\
\psi _{1}\left( r\right) &=&K_{2-2\mu }\left( z\right) ,\;\psi _{2}\left(
r\right) =K_{2\mu }\left( z\right) ,\;\frac{1}{2}<\mu <1;\,\;z=2\sqrt{\left|
1-2\mu \right| \gamma r}\,.  \label{3.7.24}
\end{eqnarray}
Quadratically integrable solutions exist for $a^{2}<1$ as well. For example,
for $0<\mu <\frac{1}{2}$\ they have the form (\ref{3.7.21}), where $c_{2}=0.$
In such a case $a^{2}$ is quantized 
\begin{equation}
a^{2}=1-\frac{\left( 1-2\mu \right) ^{2}}{\left( 1-2\mu +2n\right) ^{2}}%
\,,\;n=0,1,2,...\,.  \label{3.7.23}
\end{equation}
\ Moreover, for any complex $a^{2}\;($provided $%
\mathop{\rm Re}%
\sqrt{1-a^{2}}>0)$ there exist unbounded solutions with a finite norm. They
read 
\begin{equation}
\psi _{1}\left( r\right) =a\psi _{\lambda ,2\left( 1-\mu \right) }\left(
x\right) \,,\;\psi _{2}\left( r\right) =\left( 1+\sqrt{1-a^{2}}\right) \psi
_{\lambda ,2\mu }\left( x\right) \,,\;\lambda =\frac{1-2\mu }{2\sqrt{1-a^{2}}%
}\,\,.  \label{3.7.22}
\end{equation}

All the above solutions obey (\ref{3.7.17}) for $n=0.$

Finally we present solutions, which do not have an analog in the
Klein-Gordon case discussed in Sect. II.F. These solutions are not
eigenvectors of the radial momentum operator. With this aim in view we
present Dirac wave functions in the following form 
\begin{equation}
\Psi \left( x\right) =\Psi _{\left( -\right) }\left( x\right) +\Psi _{\left(
+\right) }\left( x\right) \,,\;\Psi _{\left( \pm \right) }\left( x\right)
=P_{\left( \pm \right) }\Psi \left( x\right) \,,\;2P_{\left( \pm \right)
}=1\pm \left( {\bf \alpha n}\right) \,,  \label{3.7.25}
\end{equation}
where ${\bf n}$ is a unit vector, ${\bf \alpha =}\left( \alpha _{k}=\gamma
^{0}\gamma ^{k}\right) ,$\ $k=1,2,3,$ and $P_{\left( \pm \right) }$ are
projection operators,$\;P_{\left( +\right) }+P_{\left( -\right)
}=1\,,\;P_{\left( \pm \right) }^{2}=P_{\left( \pm \right) }\,,\;P_{\left(
+\right) }P_{\left( -\right) }=P_{\left( -\right) }P_{\left( +\right) }=0\,.$
Then we can always present $\Psi _{\left( \pm \right) }\left( x\right) $ in
the following block form 
\begin{equation}
\Psi _{\left( +\right) }\left( x\right) =\left( 
\begin{array}{c}
\upsilon \left( x\right) \\ 
\left( {\bf \sigma n}\right) \upsilon \left( x\right)
\end{array}
\right) \,,\;\Psi _{\left( -\right) }\left( x\right) =\left( 
\begin{array}{c}
u\left( x\right) \\ 
-\left( {\bf \sigma n}\right) u\left( x\right)
\end{array}
\right) \,,  \label{3.7.27}
\end{equation}
with $u\left( x\right) ,$ $\upsilon \left( x\right) $ being arbitrary
spinors. Without loss of generality we can always chose ${\bf n=}\left(
0,0,1\right) .$ The Dirac equation (\ref{3.7.1}) demands $u\left( x\right) ,$
$\upsilon \left( x\right) $ to obey the equations \ 
\begin{eqnarray}
2\tilde{P}_{0}u &=&\left[ m_{0}c-\left( {\bf \sigma B}\right) \sigma _{3}%
\right] \upsilon \,,\;2\tilde{P}_{3}\upsilon =\left[ m_{0}c+\left( {\bf %
\sigma B}\right) \sigma _{3}\right] u\,,  \nonumber \\
2\tilde{P}_{0} &=&P_{0}-P_{3}\,,\;2\tilde{P}_{3}=P_{0}+P_{3}\,,\;{\bf B}%
=\left( P_{1},P_{2},0\right) \,.  \label{3.7.28}
\end{eqnarray}
Suppose we consider external fields, for which the operator $\tilde{P}_{3}$ (%
\ref{3.6.4}) is an integral of motion, and suppose we are looking for
solutions that are eigenvectors of the latter operator. Then in accordance
with (\ref{3.6.5}) 
\begin{equation}
2\tilde{P}_{3}=\hbar \left( \lambda +g\right) \,,\;c\hbar g=\left| e\right|
\left( A_{0}^{\left( 1\right) }+A_{3}^{\left( 1\right) }\right) \,.
\label{3.7.29}
\end{equation}
It follows from (\ref{3.7.28}) that the spinor $\upsilon $ can be restored
by the one $u,$ 
\begin{equation}
\hbar \left( \lambda +g\right) \upsilon =\left[ m_{0}c+\left( {\bf \sigma B}%
\right) \sigma _{3}\right] u\,.  \label{3.7.30}
\end{equation}
For external fields under consideration, the operator ${\bf B}$ commutes
with $\lambda +g,$ thus we get a closed equation for $u,$ 
\begin{equation}
2\hbar \left( \lambda +g\right) \tilde{P}_{0}u=\left[ m_{0}c-\left( {\bf %
\sigma B}\right) \sigma _{3}\right] \left[ m_{0}c+\left( {\bf \sigma B}%
\right) \sigma _{3}\right] u\,.  \label{3.7.31}
\end{equation}
Considering eigenvectors for the operator $J_{z}$ (\ref{3.7.8}) in
axial-symmetric external fields (\ref{3.3}), we can write 
\begin{equation}
u\left( x\right) =\left( 
\begin{array}{c}
e^{-i\varphi }u_{1}\left( r,t\right) \\ 
u_{-1}(r,t)
\end{array}
\right) \exp i\left[ \left( l-l_{0}\right) \varphi -\frac{m^{2}}{2}\int 
\frac{du^{0}}{\lambda +g\left( u^{0}\right) }-\frac{\lambda u^{3}}{2}\right]
\,,  \label{3.7.32}
\end{equation}
where the functions $u_{\zeta }\left( r,t\right) ,$ $\zeta =\pm 1$ obey the
equations 
\begin{equation}
\hat{R}_{1}^{\zeta }u_{\zeta }(r,t)=0\,,\;\hat{R}_{1}^{\zeta }=i\partial
_{t}+\partial _{r}^{2}+\frac{\partial _{r}}{r}-\frac{\left[ \bar{l}-\frac{%
1+\zeta }{2}+A\left( r\right) \right] ^{2}}{r^{2}}-\zeta \frac{A^{\prime
}\left( r\right) }{r}\,,  \label{3.7.33}
\end{equation}
which can be solved similar to the one (\ref{3.6.8}).

\section{Superposition of Aharonov-Bohm, longitudinal, and crossed fields}

We consider here Klein-Gordon and Dirac equations in some superpositions of
AB field, longitudinal, and crossed fields. In fact, there are only two
types of such fields, which admit exact solutions of these equations.

To define the first type of the fields, we introduce curvilinear coordinates 
$u^{\mu }$ by the relations 
\begin{equation}
u^{0}=x^{0}-x^{3},\;u^{1}=q\left( u^{0}\right) r^{2},\;u^{2}=\varphi
,\;u^{3}=x^{0}+x^{3}-u^{0}u^{1},\;q\left( u^{0}\right) =\left[ \left(
u^{0}\right) ^{2}+a\right] ^{-1},  \label{4a.1}
\end{equation}
where $a$ is a constant. In these coordinates, covariant components $A_{\mu
}^{\left( 1\right) }$\ of electromagnetic potentials are given as 
\begin{eqnarray}
\frac{|e|A_{0}^{\left( 1\right) }}{c\hbar } &=&q\left( u^{0}\right) \left[
f_{1}\left( u^{1}\right) +au^{1}g_{1}\left( u^{0}\right) \right]
,\;A_{1}^{\left( 1\right) }=0\,,\;  \nonumber \\
\frac{|e|A_{2}^{\left( 1\right) }}{c\hbar } &=&f_{2}\left( u^{1}\right)
+u^{1}g_{2}\left( u^{0}\right) \,,\;\frac{|e|A_{3}^{\left( 1\right) }}{%
c\hbar }=\frac{g_{1}\left( u^{0}\right) }{2}\,.  \label{4a.2}
\end{eqnarray}
Here $g_{s}\left( u^{0}\right) ,\,f_{s}\left( u^{1}\right) $ $\left(
s=1,2\right) $ are arbitrary functions of indicated arguments. The
corresponding additional (to AB field) electromagnetic field is given by its
components in cylindrical reference frame 
\begin{eqnarray}
&&\frac{|e|E_{r}}{c\hbar }=\frac{|e|H_{\varphi }}{c\hbar }=qr\left[ 2q\left(
f_{1}^{\prime }+ag_{1}\right) +u^{0}g_{1}^{\prime }\right] ,\;\frac{|e|E_{z}%
}{c\hbar }=-g_{1}^{\prime },  \nonumber \\
&&\frac{|e|E_{\varphi }}{c\hbar }=-\frac{|e|H_{r}}{c\hbar }%
=-qr[g_{2}^{\prime }-2qu^{0}\left( g_{2}+f_{2}^{\prime }\right) ],\;\frac{%
|e|H_{z}}{c\hbar }=2q\left( g_{2}+f_{2}^{\prime }\right) .  \label{4a.3}
\end{eqnarray}
Exact solutions in the field (\ref{4a.3}) were studied in \cite
{BagBiG75,BagGi90}.

In the case under considerations, the operators $L_{z}$ (or $J_{z}$ in the
Dirac equation case) and $\tilde{P}_{3}$ \ (\ref{3.6.4}) are integrals of
motion. We are going to study solutions that are eigenvectors for such
operators. Let us impose the following constraint on the functions $%
g_{s}\left( u^{0}\right) ,\,f_{s}\left( u^{1}\right) $ $\left( s=1,2\right) $
\begin{equation}
u^{1}\left( g_{2}^{2}-ag_{1}^{2}-b\right) +2g_{2}f_{2}-2g_{1}f_{1}=0,\;b=%
{\rm const\;}.  \label{4a.4}
\end{equation}
Then we can separate the variables $u^{0}\;$and $u^{1}$ and present
Klein-Gordon wave functions in the form 
\begin{eqnarray}
\Psi &=&\sqrt{\frac{q}{P}}e^{-i\Gamma }\psi \left( u^{1}\right) ,\;P=\lambda
+g_{1}\left( u^{1}\right) \,,  \nonumber \\
\,\Gamma &=&\frac{\lambda }{2}u^{3}-\left( l-l_{0}\right) \varphi +\int %
\left[ m^{2}+2q\left( 2k_{1}+\bar{l}g_{2}\right) \right] \frac{du^{0}}{2P}\,.
\label{4a.5}
\end{eqnarray}
The functions $\psi \left( u^{1}\right) $ obey the equation 
\begin{equation}
\psi ^{\prime \prime }+\frac{1}{u^{1}}\psi ^{\prime }+R\left( u^{1}\right)
\psi =0\,,\;R\left( u^{1}\right) =\frac{2k_{1}+\lambda f_{1}}{2u^{1}}-\frac{%
a\lambda ^{2}+b}{4}-\frac{\left( \bar{l}+f_{2}\right) }{4\left( u^{1}\right)
^{2}}^{2}\,.  \label{4a.6}
\end{equation}
In the same case, Dirac wave functions have the form 
\begin{equation}
\Psi =\frac{\sqrt{q}}{P}e^{-i\Gamma }KW\left[ \left( 1+\sigma _{3}\right)
\psi _{1}\left( u^{1}\right) +\left( 1-\sigma _{3}\right) \psi _{-1}\left(
u^{1}\right) \right] \upsilon \,,  \label{4a.8}
\end{equation}
where $\upsilon $ is an arbitrary constant spinor, and 
\begin{eqnarray}
&&K=\left( 
\begin{array}{c}
m+P-\sigma _{3}\left( {\bf \sigma F}\right) \\ 
\left( m-P\right) \,\,\sigma _{3}-\left( {\bf \sigma F}\right)
\end{array}
\right) \,,\;{\bf F}={\bf e}_{r}\,qr\left( 2i\partial _{u^{1}}-u^{0}P\right)
\nonumber \\
&&+{\bf e}_{\varphi }\left( \frac{\bar{l}+f_{2}}{r}+qrg_{2}\right)
\,,\;W=\cos \delta +i\sigma _{3}\sin \delta ,\;\delta =\int \frac{qg_{2}}{P}%
du^{0}.  \label{4a.9}
\end{eqnarray}
The scalar functions $\psi _{\zeta }\left( u^{1}\right) \;\left( \zeta =\pm
1\right) \,\,$obey a set of independent equations 
\begin{equation}
\psi _{\zeta }^{\prime \prime }+\frac{1}{u^{1}}\psi _{\zeta }^{\prime }+%
\left[ R\left( u^{1}\right) +\zeta \frac{f\,\,_{2}^{\prime }}{2u^{1}}\right]
\psi _{\zeta }=0\,\,.  \label{4a.12}
\end{equation}
Explicit solutions of the equations (\ref{4a.6}) and (\ref{4a.12}) can be
written for 
\begin{equation}
f_{1}\left( u^{1}\right) =\alpha u^{1}+\frac{\beta }{u^{1}}%
\,,\;\,\,f_{2}\left( u^{1}\right) =\gamma u^{1},\;\,\alpha ,\,\beta
,\,\gamma ={\rm const\,}.  \label{4a.13}
\end{equation}
In such a case the equations (\ref{4a.6}) and (\ref{4a.12}) are reduced to
the one (\ref{3.5.1}). Solutions of the latter equation we have studied
above.

Let us return to the constraint (\ref{4a.4}). If $\beta \neq 0,$ then $%
g_{1}=0,\;g_{2}=$ ${\rm const},$ and $b$ can be found from (\ref{4a.4}) to
be $b=g_{2}^{2}+2\gamma g_{2}\,.$ If $\beta =0,$ then $g_{1},\;g_{2}$ are
related by an equation 
\begin{equation}
\left( g_{2}+\alpha \right) ^{2}=a\left( g_{1}+\frac{\gamma }{a}\right)
^{2}+\alpha ^{2}-\frac{\gamma ^{2}}{a}+b\,.  \label{4a.15}
\end{equation}
Thus, one of the constants remains arbitrary. We see that there exist a wide
class of fields, which admit exact solutions.

To define the second type of fields, which admit exact solutions, we
introduce curvilinear coordinates $u^{\mu }$ by the relations 
\begin{equation}
u^{0}=x^{0}-x^{3},\;u^{1}=\frac{r^{2}}{u^{0}},\;u^{2}=\varphi
,\,\,u^{3}=x^{0}+x^{3}-\frac{r^{2}}{2u^{0}}\,.  \label{4a.16}
\end{equation}
Covariant components of electromagnetic potentials in the coordinates (\ref
{4a.16}) are given by the equations\bigskip 
\begin{equation}
\frac{\left| e\right| A_{0}^{\left( 1\right) }}{c\hbar }=\frac{f_{1}\left(
u^{1}\right) }{u^{0}},\;\,A_{1}^{\left( 1\right) }=0,\;\frac{\left| e\right|
A_{2}^{\left( 1\right) }}{c\hbar }=f_{2}\left( u^{1}\right)
,\;\,A_{3}^{\left( 1\right) }=0\,,  \label{4a.17}
\end{equation}
where $\,f_{s}\left( u^{1}\right) $ $\left( s=1,2\right) $ are arbitrary
functions of $u^{1}.$ The corresponding electromagnetic field is given by
its components in the cylindrical reference frame\bigskip 
\begin{equation}
E_{r}=H_{\varphi }=\frac{2c\hbar r}{\left| e\right| \left( u^{0}\right) ^{2}}%
f\,_{1}^{\,\prime }\,,\,\,E_{\varphi }=-H_{r}=-\frac{2c\hbar r}{\left|
e\right| \left( u^{0}\right) ^{2}}f\,_{2}^{\,\prime }\,,\,\,H_{z}=\frac{%
2c\hbar }{\left| e\right| u^{o}}\,f\,\,_{2}^{\prime }\,,\;\,E_{z}=0\,.
\label{4a.18}
\end{equation}
In the absence of AB field, exact solutions in such a field were studied in 
\cite{BagGiZ78,BagGi90}.

Here integrals of motion are the same as in the previous case. Klein-Gordon
wave functions can be written in the form 
\begin{equation}
\Psi =\frac{1}{\sqrt{u^{0}}}^{\,}\psi \left( u^{1}\right) \exp \left\{ -i%
\left[ \frac{\lambda }{2}u^{3}-i\left( l-l_{0}\right) \varphi
+m^{2}u^{0}+k_{1}\ln u^{0}\right] \right\} \,\,.  \label{4a.19}
\end{equation}
The functions $\psi \left( u^{1}\right) $ obey the equation 
\begin{equation}
\psi ^{\prime \prime }+\frac{1}{u^{1}}\psi ^{\prime }+R\left( u^{1}\right)
\psi =0\,,\;R\left( u^{1}\right) =\frac{\lambda ^{2}}{16}+\frac{%
k_{1}+2\lambda f_{1}}{4u^{1}}-\frac{\left( \bar{l}+f_{2}\right) ^{2}}{%
4\left( u^{1}\right) ^{2}}\,\,.  \label{4a.20}
\end{equation}
Dirac wave function can be presented in the form (\ref{4a.8}), (\ref{4a.9})
with the following modifications 
\begin{equation}
P=\lambda \,,\,\,q=\frac{1}{u^{0}}\,,\,{\bf F}={\bf e}_{r}\,\,\frac{r}{u^{0}}%
\left( 2i\partial _{1}-\frac{\lambda }{2}\right) +{\bf e}_{\varphi }\frac{%
\bar{l}+f_{2}}{r}\,\,.  \label{4a.21}
\end{equation}
Besides, the functions $\psi _{\zeta }\left( u^{1}\right) $ have to obey the
equation (\ref{4a.12}) with $R\left( u^{1}\right) $ defined by (\ref{4a.20}%
). Solutions of the latter equations are available for $f_{s}\left(
u^{1}\right) $ in the form (\ref{4a.13}). Thus, we have again returned to
the equation (\ref{3.5.1}).

\section{Superposition of Aharonov-Bohm field and some non-uniform fields}

Consider now additional fields, which are given by potentials of the form 
\begin{equation}
A_{0}^{\left( 1\right) }=\frac{c\hbar }{\left| e\right| }f_{1}\left(
r\right) \,,\,\;\,A_{1}^{\left( 1\right) }\,\,=\frac{c\hbar }{\left|
e\right| }\frac{A\left( r\right) }{r}\sin \varphi \,,\;\,A_{2}^{\left(
1\right) }=-\frac{c\hbar }{\left| e\right| }\frac{A\left( r\right) }{r}\cos
\varphi \,,\;\,A_{3}^{\left( 1\right) }=\frac{c\hbar }{\left| e\right| }%
\,f_{2}\left( r\right) \,.  \label{5.1}
\end{equation}
Here $f_{1}\left( r\right) \,,\,\,f_{2}\left( r\right) ,\;A\left( r\right) $
are arbitrary functions of $r.$ The corresponding electromagnetic field
components in cylindrical reference frame have the form 
\begin{equation}
E_{r}=-\frac{c\hbar }{\left| e\right| }f\,_{1}^{\prime }\left( r\right)
,\,H_{\varphi }=\frac{c\hbar }{\left| e\right| }f_{2}^{\,\,\,\prime }\left(
r\right) ,\,\,H_{z}=\frac{c\hbar }{\left| e\right| }\frac{A^{\prime }\left(
r\right) }{r}\,,\;E_{\varphi }=E_{z}=H_{r}=0\,.  \label{5.2}
\end{equation}
Exact solutions of relativistic wave equations in such fields were studied
in \cite{BagGiZ80,BagGi90}. Here we present exact solutions of the equations
in the superposition of these fields and AB field.

Stationary solutions of Klein-Gordon equation that are eigenvectors for the
operators $p_{0},p_{3}$, $L_{z}$ can be written as

\begin{equation}
\Psi \left( x\right) =e^{-i\Gamma }\psi \left( r\right) \,,\,\,\Gamma
=\,k_{0}x^{o}+k_{3}x^{3}-\,\left( l-l_{0}\right) \varphi \,.  \label{5.3}
\end{equation}
Functions $\psi \left( r\right) $ obey the equation 
\begin{equation}
\psi ^{\prime \prime }\left( r\right) +\frac{1}{r}\psi ^{\prime }\left(
r\right) +R\left( r\right) \psi \left( r\right) +0\,,\;R\left( r\right)
=\left( k_{0}+f_{1}\right) ^{2}-\left( k_{3}+f_{2}\right) ^{2}-\frac{\left( 
\bar{l}+A\right) ^{2}}{r^{2}}-m^{2}\,.  \label{5.4}
\end{equation}
The corresponding solutions of Dirac equation have the form $\Psi \left(
x\right) =e^{-i\Gamma }M\psi ,$ where the matrix $M$ reads $M={\rm diag}%
\left( e^{-i\varphi },i,e^{-i\varphi },i\right) ,$ and the bispinor $\psi $=$%
(\psi _{k})\;\left( k=1,2,3,4\right) $ obeys the equation 
\begin{equation}
\left[ k_{0}+f_{1}-m\gamma ^{0}-\frac{1}{r}\left( \bar{l}+A-\frac{1}{2}%
\right) \alpha _{1}-i\left( \frac{d}{dr}+\frac{1}{2r}\right) \alpha
_{2}+\left( k_{3}+f_{2}\right) \alpha \right] \psi =0\,.  \label{5.6}
\end{equation}
In some particular cases the latter equation can be reduced to the one (\ref
{3.7.10}) and thus solved explicitly. All such cases are described in \cite
{BagGiZ80,BagGi90}.

\section{Aharonov-Bohm field in 2+1 QED}

Consider Dirac equation in $2+1$ dimensions (see for example, \cite
{JacNa91,GitSh97}) ($x=(x^{\mu }),\,\,\mu =0,1,2,\;\gamma ^{0}=\sigma
_{3},\,\,\gamma ^{1}=i\sigma _{2},\,\,\gamma ^{3}=-i\sigma _{1}$), 
\begin{equation}
(\gamma ^{\mu }P_{\mu }-m_{0}c)\,\Psi \left( x\right) =0\,,\;\,\Psi \left(
x\right) =\left( 
\begin{array}{c}
\Psi _{1}\left( x\right) \\ 
\Psi _{2}\left( x\right)
\end{array}
\right) \,\,.  \label{6.1}
\end{equation}
For components of the spinor $\Psi \left( x\right) $ we get the following
equations 
\begin{equation}
\left( P_{0}-m\right) \Psi _{1}+\left( P_{1}-iP_{2}\right) \Psi
_{2}=0\,\,,\;\left( P_{0}+m\right) \Psi _{2}+\left( P_{1}+iP_{2}\right) \Psi
_{1}=0\,.  \label{6.3}
\end{equation}
These equations can be solved exactly for a superposition of AB field and an
additional field described below. Potentials (\ref{2.10}) of the latter
field are given as 
\begin{equation}
\frac{\left| e\right| }{c\hbar }A_{0}^{\left( 1\right) }=B\left( r\right) ,\,%
\frac{\left| e\right| }{c\hbar }\,\,A_{1}^{\left( 1\right) }=\frac{A\left(
r\right) }{r}\sin \varphi ,\,\,\frac{\left| e\right| }{c\hbar }A_{2}^{\left(
1\right) }=-\frac{A\left( r\right) }{r}\cos \varphi \,\,,  \label{6.4}
\end{equation}
where $A\left( r\right) ,B\left( r\right) \ $are arbitrary functions of $r.$
This field is an analog of the field (\ref{5.1}) in $3+1$ dimensions.
Potentials $A_{0}^{\left( 0\right) },\,A_{1}^{\left( 0\right)
},\,A_{2}^{\left( 0\right) }\;$of AB field in $2+1$ dimensions are still
given by the formulas (\ref{2.1}). The operators 
\begin{equation}
p_{0}=i\hbar \frac{\partial }{\partial \,x^{0}},\,\,\,J_{3}=-i\hbar \frac{%
\partial }{\partial \varphi }+\frac{\hbar }{2}\sigma _{3}  \label{6.5}
\end{equation}
are integrals of motions in the external field under consideration. Thus, we
can impose the following conditions on the spinor $\Psi $ 
\begin{equation}
p_{0}\Psi =\hbar k_{0}\Psi ,\,\,\,J_{3}\Psi =\hbar \left( l-l_{0}-\frac{1}{2}%
\right) \Psi \,.  \label{6.6}
\end{equation}
A solution of the equations (\ref{6.1}), (\ref{6.6}) has the form 
\begin{equation}
\Psi \left( x\right) =e^{-i\Gamma }\left( 
\begin{array}{c}
e^{-i\varphi }\psi _{1}\left( r\right) \\ 
i\psi _{2}\left( r\right)
\end{array}
\right) \,,\,\,\Gamma =k_{0}x^{0}-\left( l-l_{0}\right) \varphi \,,
\label{6.7}
\end{equation}
where the functions $\psi _{k}\left( r\right) $ $\left( k=1,2\right) $ obey
the equations 
\begin{eqnarray}
\psi _{1}^{\prime }\left( r\right) &=&\frac{\bar{l}-1+A\left( r\right) }{r}%
\psi _{1}\left( r\right) -\left( k_{0}+B\left( r\right) +m\right) \psi
_{2}\left( r\right) \,,  \nonumber \\
\psi _{2}^{\prime }\left( r\right) &=&\left( k_{0}+B\left( r\right)
-m\right) \psi _{1}\left( r\right) -\frac{\bar{l}+A\left( r\right) }{r}\psi
_{2}\left( r\right) \,\ .  \label{6.8}
\end{eqnarray}
Explicit solutions of these equations can be found in three particular
cases: 
\begin{eqnarray}
1.\,\,\,A\left( r\right) &=&B\left( r\right) =0\,;  \label{6.9} \\
2.\,\ A\left( r\right) &=&\rho =\frac{\gamma r^{2}}{2},\,\,\,\gamma
>0,\;B\left( r\right) =0\,;  \label{6.10} \\
3.\,\,A\left( r\right) &=&\gamma r,\,\,B\left( r\right) =\frac{b}{r}%
,\,\,\gamma >0\,;  \label{6.11}
\end{eqnarray}
Below we consider each case in detail.

The case (\ref{6.9}) corresponds to the pure AB field. For $l\neq 0,$ there
exist bounded solutions of the form 
\begin{eqnarray}
\psi _{1}\left( r\right) &=&\sqrt{k_{0}+m\,\,\,}J_{\left| \bar{e}-1\right|
}\left( kr\right) ,\,\,\,k=\sqrt{k_{0}^{2}-m^{2}}\,\,,  \nonumber \\
\psi _{2}\left( r\right) &=&\varepsilon \sqrt{k_{0}+m\,\,\,}J_{\left| \bar{e}%
\right| }\left( kr\right) \,\,,\,\ \varepsilon ={\rm sign\,}l\,\,.
\label{6.12}
\end{eqnarray}
Here $J_{l}(x)$ are Bessel functions. Solutions (\ref{6.12}) vanish at $r=0.$
They are orthogonal and normalized.

For $l=0,$ $\mu \neq 0,$ bounded solutions do not exist. However, there are
unbounded at $r=0$ solutions. Some of them have the form 
\begin{eqnarray}
&&\psi _{1}\left( r\right) =\sqrt{k_{0}+m}\left[ c_{1}J_{\mu -1}\left(
kr\right) +c_{2}J_{1-\mu }\left( kr\right) \right] \,\,,  \nonumber \\
&&\psi \left( r\right) =\sqrt{k_{0}-m}\left[ c_{1}J_{\mu }\left( kr\right)
-c_{2}J_{-\mu }\left( kr\right) \right] \,.  \label{6.13}
\end{eqnarray}
Here $c_{1},c_{2}$ are arbitrary constants. The solutions (\ref{6.13}) are
still orthogonal and normalized. Another set of unbounded solutions (they
are expressed via the Macdonald functions) reads 
\begin{eqnarray}
\psi _{1}\left( r\right) &=&\sqrt{m+k_{0}}K_{1-\mu }\left( \sqrt{%
m^{2}-k_{0}^{2}}r\right) ,\,\,%
\mathop{\rm Re}%
\,\sqrt{m^{2}-k_{0}^{2}}>0\,\,,  \nonumber \\
\psi _{2}\left( r\right) &=&-\sqrt{m-k_{0}}K_{\mu }\left( \sqrt{%
m^{2}-k_{0}^{2}}r\right) .\,  \label{6.14}
\end{eqnarray}
As we see they are defined even for some complex $k_{0}.$ It is interesting
to remark that the latter solutions have finite norms.

The case (\ref{6.10}) corresponds a combination of AB and uniform constant
magnetic fields. Here we can introduce operators $a_{k}^{+},a_{k}$ $\left(
k=1,2\right) $ by relations (\ref{3.4.1}). Using the substitution

\begin{equation}
\psi _{1}\left( x\right) =e^{-ik_{0}x^{0}}\,\psi _{1}\left( \rho ,\varphi
\right) \,,\;\psi _{2}\left( x\right) =ie^{-ik_{0}x^{0}}\,\psi _{2}\left(
\rho ,\varphi \right) \,,  \label{6.15}
\end{equation}
we present (\ref{6.3}) in the following form 
\begin{equation}
\left( k_{0}-m\right) \psi _{1}\left( \rho ,\varphi \right) -\sqrt{2\gamma }%
a_{1}\psi _{2}\left( \rho ,\varphi \right) =0\,\,,\;\left( k_{0}+m\right)
\psi _{2}\left( \rho ,\varphi \right) -\sqrt{2\gamma }a_{1}^{+}\psi
_{1}\left( \rho ,\varphi \right) =0\,\,.  \label{6.16}
\end{equation}
Now we can use the functions (\ref{3.4.15}) and the relations (\ref{3.4.9}).

Consider first states with $k_{0}^{2}\neq m^{2},\;l\neq 0.$ As in $3+1$
dimensions, these states can be divided in two types ( $j=1,2$), 
\begin{equation}
\psi _{1}\left( \rho ,\varphi \right) =\sqrt{k_{0}+m}\psi
_{n-1,\,\,\,l-1}^{\left( j\right) }\,,\,\psi _{2}\left( \rho ,\varphi
\right) =\sqrt{k_{0}-m}\psi _{n,l}^{\left( j\right)
}\,\,,\;k_{0}^{2}=m^{2}+2\gamma \bar{n}\,\,,  \label{6.17}
\end{equation}
where $\bar{n}$ was defined in (\ref{3.4.14}).\ \ Solutions (\ref{6.17})
vanish at $r=0.$

If $k_{0}^{2}\neq m^{2},\;l=0,$ $\mu \neq 0,$ then solutions, which are
formally defined by equations (\ref{6.17}), are unbounded at $r=0.$ However,
they are still orthogonal and normalized. For$\;l=0,$ $\mu \neq 0,$ $%
k_{0}^{2}=m^{2}+\gamma \left( 2\lambda +\mu -1\right) $ ($\lambda $ is
arbitrary complex), there exist another unbounded solutions. They have the
form (\ref{6.7}), with 
\begin{equation}
\psi _{1}\left( r\right) =\left( k_{0}+m\right) \psi _{\lambda -\frac{1}{2}%
,\,1-\mu }\left( \rho \right) ,\,\psi _{2}\left( r\right) =\sqrt{2\gamma }%
\psi _{\lambda ,\mu }\left( \rho \right) \,,  \label{6.18}
\end{equation}
where the functions $\psi _{\lambda ,\mu }\left( x\right) $ are defined by (%
\ref{4.139}), (\ref{4.140}). These solutions are orthogonal and normalized
as well.

Consider now states with $k_{0}^{2}=m^{2}.$ Suppose $k_{0}=m;$ then a
general solutions of Eqs. (\ref{6.8}) reads 
\begin{eqnarray}
\psi _{1}\left( r\right) &=&Nf_{l}\left( \rho \right) +ce^{\frac{\rho }{2}%
}\rho ^{\frac{\bar{l}-1}{2}}\,,\,\;\psi _{2}\left( r\right) =gNe^{-\frac{%
\rho }{2}}\rho ^{-\frac{\bar{l}}{2}}\,\ ,  \nonumber \\
f_{l}\left( \rho \right) &=&e^{\frac{\rho }{2}}\,\rho ^{\frac{\bar{l}-1}{2}%
}\,\,\stackrel{\infty }{%
\mathrel{\mathop{\int }\limits_{\rho }}%
}e\,^{-x}x^{-\bar{l}}\,dx,\,\,\ g=\sqrt{\frac{\gamma }{2m^{2}}}\,\,,
\label{6.19}
\end{eqnarray}
where $N,\,c$ are arbitrary constants. For $\mu \neq 0$ the only some states
with $l=0$ have a finite norm. Namely the states $\ $ 
\begin{eqnarray}
&&\psi _{1}\left( r\right) =N\rho ^{\frac{\mu -1}{2}}e^{\frac{\rho }{2}}\,%
\stackrel{\infty }{%
\mathrel{\mathop{\int }\limits_{\rho }}%
}\,e^{-x}x^{-\mu }dx,\,\,\psi _{2}\left( r\right) =gN\,\,e^{-\frac{\rho }{2}%
}\rho ^{-\frac{\mu }{2}}\,\,,  \nonumber \\
&&N=\sqrt{\frac{\Gamma \left( 1+\mu \right) \sin \mu \pi }{\pi \left[ 1+\mu
\psi \left( 1\right) -\mu \psi \left( 1+\mu \right) +\mu g^{2}\right] }}\;,\;%
\stackrel{\infty }{%
\mathrel{\mathop{\int }\limits_{0}}%
}\,\left[ \psi _{1}^{2}\left( r\right) +\psi _{2}^{2}\right] d\rho =1\,,
\label{6.20}
\end{eqnarray}
where $\psi \left( x\right) $ is the logarithmic derivative of $\Gamma -$%
function (\cite{GraRy94}, 8.360). All the above solutions are singular at $%
r=0.$ One can see that $\lim_{\mu \rightarrow 0}\,N=0.$ Thus, in $2+1$
dimensions, in pure magnetic field, Dirac equation does not have solutions
with $k_{0}=m,$ in contrast to the corresponding $3+1-$dimensional case. For 
$\mu \neq 0$ (in the presence of AB field) such solutions appear even in $%
2+1 $ dimensions.

Suppose $k_{0}=-m.$ Then, the functions 
\begin{equation}
\psi _{1}\left( r\right) =0,\,\,\psi _{2}\left( r\right) =\left[ \Gamma
\left( 1-\bar{l}\right) \right] ^{-\frac{1}{2}\ \,}\rho ^{-\frac{\bar{l}}{2}%
}\,e^{-\frac{\rho }{2}}\,,\;\stackrel{\infty }{%
\mathrel{\mathop{\int }\limits_{0}}%
}\,\left[ \psi _{1}^{2}\left( r\right) +\psi _{2}^{2}\right] d\rho =1\,
\label{6.24}
\end{equation}
present solutions for $l\leq 0.$ One can see that (\ref{6.24}) is a
particular case of (\ref{6.17}). For $l<0,$ the states (\ref{6.24}) vanish
at $r=0.$ For $l=0,$ $\mu \neq 0,$ they are singular at $r=0.$ If $\mu =0$,
then this singularity disappears, however the states do not tend to zero as $%
r\rightarrow 0.$

Let us finally turn to the case (\ref{6.11}). Here it is enough to consider $%
b\neq 0$ only. Indeed, suppose $b=0;$ then doing the change of functions ($%
\psi \rightarrow f$) 
\begin{equation}
\psi _{1}\left( r\right) =\sqrt{k_{0}+m}\,f_{1}\left( r\right) \,,\,\,\,\psi
_{2}\left( r\right) =\sqrt{k_{0}-m}\,f_{2}\left( r\right) \,\,,  \label{6.25}
\end{equation}
we can transform Eqs. (\ref{6.8}) to the form (\ref{3.7.10}) with $k_{1}=%
\sqrt{k_{0}^{2}-m^{2}}\,.$ Solutions of Eqs. (\ref{3.7.10}) are given by the
formulas (\ref{3.7.18})-(\ref{3.7.24}). Thus, below we consider the case $%
b\neq 0$ only. We introduce the following notation 
\begin{equation}
g_{1}=\sqrt{|2\bar{l}-1|+2b}\,,\;\,g_{-1}=\sqrt{|2\bar{l}-1|-2b}%
\,,\,\,\varepsilon =\pm 1=\,{\rm \,sign\,}l\,.  \label{6.27}
\end{equation}
Thus solutions of Eqs. (\ref{6.8}) have the form 
\begin{eqnarray}
&&\psi _{1}\left( r\right) =c_{1}I_{\alpha +s,s}\left( y\right)
+c_{2}I_{\alpha +s-1,\,\,s+1}\left( y\right) \,,\;\psi _{2}\left( r\right)
=c_{3}I_{\alpha +s,s}\left( y\right) +c_{4}I_{\alpha +s-1,\,\,s+1}\left(
y\right) \,,  \nonumber \\
&&\,y=2\sqrt{m^{2}+\gamma ^{2}-k_{0}^{2}\,}\,\,r\,\,,\;\alpha
=1+g_{1}\,g_{-1\,\,},\;2+2s+g_{1}g_{-1}=\frac{2bk_{0}-\left( 2\bar{l}%
-1\right) \gamma }{\sqrt{m^{2}+\gamma ^{2}-k_{0}^{2}}}\,\,,\;  \label{6.28}
\end{eqnarray}
where 
\begin{eqnarray}
c_{1} &=&\left( g_{\varepsilon }-\varepsilon g_{-\varepsilon }\right) \sqrt{%
(2\bar{l}-1)k_{0}-2b\gamma -mg_{1}g_{-1}}=\frac{\left( g_{\varepsilon
}-\varepsilon g_{-\varepsilon }\right) }{\left( g_{\varepsilon }+\varepsilon
g_{-\varepsilon }\right) }c_{3}\,,  \nonumber \\
c_{2} &=&-\left( g_{\varepsilon }+\varepsilon g_{-\varepsilon }\right) \sqrt{%
(2\bar{l}-1)k_{0}-2b\gamma +mg_{1}g_{-1}\,}=\frac{\left( g_{\varepsilon
}+\varepsilon g_{-\varepsilon }\right) }{\left( g_{\varepsilon }-\varepsilon
g_{-\varepsilon }\right) }c_{4}\,,  \label{6.29}
\end{eqnarray}
and the relations (\ref{4.11})-(\ref{4.16}) for the Laguerre functions $%
I_{n,m}$ were used. It follows from (\ref{6.27}) that both $g_{1}$ and $%
g_{-1}$ are real and positive for $\left( 2\bar{l}-1\right) ^{2}>4b^{2}\,.$
If $\left( 2\bar{l}-1\right) ^{2}<4b^{2},$ then one of these quantities is
real and positive and another one is imaginary. Suppose $\left( 2\bar{l}%
-1\right) ^{2}>4b^{2}.$ Then for the energies $k_{0}^{2}<m^{2}+\gamma ^{2}$
there exist bound states$.$ In the latter case $s$ is positive integer and $%
k_{0}^{2}$ is quantized according to the last equation (\ref{6.28}).

Another observation: For any $b$ there exist a number $l$ such that 
\begin{equation}
1>g_{1}^{2}\,\,g_{-1}^{2}=\left( 2\bar{l}-1\right) ^{2}-4b^{2}>0\,\,.
\label{6.31}
\end{equation}
For example, $b=0$ corresponds to $l=0.$ Then one can chose two values for $%
\alpha ,$ 
\begin{equation}
\alpha _{1}=1+g_{1}g_{-1\,},\,\,\,\alpha _{2}=1-g_{1}g_{-1}\,\,.
\label{6.32}
\end{equation}
In both cases the solutions (\ref{6.28}) have singularity at $r=0$ but still
have finite norms.

\begin{acknowledgement}
V.G.B thanks FAPESP for support and Nuclear Physics Department of S\~{a}o
Paulo University for hospitality, as well as he thanks Russian Science
Ministry Foundation and RFFI for partial support; D.M.G thanks FAPESP and
CNPq for permanent support.
\end{acknowledgement}

\section{Appendix}

1.The Laguerre functions $I_{n,m}(x)$ are defined by the relation

\begin{equation}
I_{n,m}(x)=\sqrt{\frac{{\Gamma (1+n)}}{{\Gamma (1+m)}}}\,{\frac{\exp (-x/2)}{%
{\Gamma (1+n-m)}}}x^{\frac{{n-m}}{2}}\Phi (-m,n-m+1;x).\;  \label{4.1}
\end{equation}
Here $\Phi \left( \alpha ,\gamma ;x\right) $ is the confluent hypergeometric
function in a standard definition (see \cite{GraRy94}, 9.210). For $\gamma
\neq -s,$ where $s$ is integer and non-negative, the latter function can be
presented by a series 
\begin{equation}
\Phi (\alpha ,\gamma ;x)=\sum_{k=0}^{\infty }{\frac{{(\alpha )_{k}}}{{%
(\gamma )_{k}}}}{\frac{x^{k}}{k!}}={\frac{\Gamma (\gamma )}{\Gamma (\alpha )}%
}\sum_{k=0}^{\infty }{\frac{{\Gamma (\alpha +k)}}{{\Gamma (\gamma +k)}}}{%
\frac{x^{k}}{k!}}\;.  \label{4.2}
\end{equation}
This series is converges for any complex $x$. For any complex $\alpha ,$ the
Pochhammer symbols ${(\alpha )_{k}}$ are defined as follows 
\begin{equation}
(\alpha )_{k}=\alpha (\alpha +1)...(\alpha +k-1)={\frac{\Gamma (\alpha +k)}{%
\Gamma (\alpha )}}\;.  \label{4.3}
\end{equation}

2. Let $m$ be a non-negative integer number; then the Laguerre functions are
related to Laguerre polynomials $L_{n}^{\alpha }(x)$ (\cite{GraRy94}, 8.970,
8.972.1) by the equation

\begin{equation}
I_{n,m}(x)=\sqrt{\frac{{\Gamma (1+m)}}{{\Gamma (1+n)}}}\exp (-x/2)x^{\frac{{%
n-m}}{2}}L_{m}^{n-m}(x),\;m=0,1,2,...\;,  \label{4.4}
\end{equation}

\begin{equation}
L_{n}^{\alpha }(x)={\frac{1}{n!}}e^{x}x^{-\alpha }{\frac{{d^{n}}}{{dx^{n}}}}%
e^{-x}x^{n+\alpha }=\sum_{k=0}^{n}\left( 
\begin{array}{c}
n+\alpha \\ 
n-k
\end{array}
\right) {\frac{(-x)^{k}}{k!}}=\left( 
\begin{array}{c}
n+\alpha \\ 
n
\end{array}
\right) \Phi (-n,1+\alpha ;x)\;.  \label{4.5}
\end{equation}
Here $n$ are non-negative integer numbers such that 
\[
\left( 
\begin{array}{c}
\alpha \\ 
n
\end{array}
\right) =\frac{{\Gamma (1+\alpha )}}{{\Gamma (1+n)\Gamma (1+\alpha -n)}}{=}%
\frac{{\alpha (\alpha -1)...(\alpha -n+1)}}{n!}{\,.} 
\]

3. Using well-known properties of the confluent hypergeometric function ( 
\cite{GraRy94}, 9.212; 9.213; 9.216), one can easily get both relations for
the Laguerre functions

\begin{eqnarray}
&&2\sqrt{x(n+1)}I_{n+1,m}(x)=(n-m+x)I_{n,m}(x)-2xI_{n,m}^{\prime }(x)\;,
\label{4.11} \\
&&2\sqrt{x(m+1)}I_{n,m+1}(x)=(n-m-x)I_{n,m}(x)+2xI_{n,m}^{\prime }(x)\;,
\label{4.12} \\
&&2\sqrt{xn}I_{n-1,m}(x)=(n-m+x)I_{n,m}(x)+2xI_{n,m}^{\prime }(x)\;,
\label{4.13} \\
&&2\sqrt{xm}I_{n,m-1}(x)=(n-m-x)I_{n,m}(x)-2xI_{n,m}^{\prime }(x)\;,
\label{4.14} \\
&&2\sqrt{nm}I_{n-1,m-1}(x)=(n+m-x)I_{n,m}(x)-2xI_{n,m}^{\prime }(x)\;,
\label{4.15} \\
&&2\sqrt{(n+1)(m+1)}I_{n+1,m+1}(x)=(n+m+2-x)I_{n,m}(x)  \nonumber \\
&&+2xI_{n,m}^{\prime }(x)\;,  \label{4.16}
\end{eqnarray}
and a differential equation for these functions

\begin{equation}
4x^{2}I_{n,m}^{\prime \prime }(x)+4xI_{n,m}^{\prime
}(x)-[x^{2}-2x(1+n+m)+(n-m)^{2}]I_{n,m}(x)=0\;.  \label{4.17}
\end{equation}
Suppose $I_{n,m}(x)$ and $I_{m,n}(x)$ are linearly independent. Then, a
general solution $I$ \ of this equation has the form $\
I=AI_{n,m}(x)+BI_{m,n}(x).$ However, whenever the condition (\ref{4.25})
holds, $I_{n,m}(x)$ and $I_{m,n}(x)$ are dependent. The formulas (\ref{4.11}%
)-(\ref{4.16}) and the equation (\ref{4.17}) are valid for any complex $%
n,m,x $ . One has to be careful applying the formulas (\ref{4.13})-(\ref
{4.15}) for $n,m=0$ . A straightforward calculation, which uses (\ref{4.1})
and (\ref{4.2}), gives

\begin{equation}
\lim_{n\rightarrow 0}\sqrt{n}I_{n-1,m}(x)=-{\frac{{\sin m\pi }}{\pi }}\sqrt{%
\Gamma (1+m)}x^{-{\frac{{1+m}}{2}}}\exp (x/2),\;\lim_{m\rightarrow 0}{\sqrt{m%
}I_{n,m-1}(x)}=0\;.  \label{4.20}
\end{equation}
A combination of Eqs. (\ref{4.11})-(\ref{4.14}) results in the following
relations

\begin{eqnarray}
&&2\sqrt{x}I_{n,m}^{\prime }(x)=\sqrt{n}I_{n-1,m}\left( x\right) -\sqrt{n+1}%
I_{n+1,m}(x)  \nonumber \\
&&=\sqrt{m+1}I_{n,m+1}\left( x\right) -\sqrt{m}I_{n,m-1}\left( x\right) ,
\label{4.21} \\
&&\sqrt{x\left( n+1\right) }I_{n+1,m}\left( x\right) -\left( n-m+x\right)
I_{n,m}\left( x\right) +\sqrt{xn}I_{n-1,m}\left( x\right) =0\,,  \label{4.22}
\\
&&\sqrt{x\left( m+1\right) }I_{n,m+1}\left( x\right) -\left( n-m-x\right)
I_{n,m}\left( x\right) +\sqrt{xm}I_{n,m-1}\left( x\right) =0\,.  \label{4.23}
\end{eqnarray}

4. Using properties of the confluent hypergeometric function, one can get a
representation

\begin{equation}
I_{n,m}(x)=\sqrt{\frac{{\Gamma (1+n)}}{{\Gamma (1+m)}}}{\frac{{\exp }\left(
x/2\right) }{{\Gamma (1+n-m)}}}x^{\frac{{n-m}}{2}}\Phi (1+n,1+n-m;-x)\;,
\label{4.24}
\end{equation}
and a relation (\cite{GraRy94}, 9.214)

\begin{equation}
I_{n,m}(x)=(-1)^{n-m}I_{m,n}(x),\;n-m\;{\rm integer}\;.  \label{4.25}
\end{equation}

5. An asymptotic formula takes place

\begin{equation}
\Phi (a,c;x)\approx {\frac{{\Gamma (c)}}{\Gamma (a)}}e^{x}x^{a-c}\;,\;%
\mathop{\rm Re}%
x\rightarrow \infty \,.  \label{4.28}
\end{equation}
Thus we obtain the following asymptotic behavior of $I_{n,m}(x)$ whenever $m$
is not integer

\begin{equation}
I_{n,m}(x)={-{\frac{{\sin m\pi }}{\pi }}}\sqrt{\Gamma (1+n)\Gamma (1+m)}x^{{-%
{\frac{{n+m+2}}{2}}}}\exp (x/2),\;%
\mathop{\rm Re}%
x\rightarrow \infty \,,  \label{4.29}
\end{equation}
and

\begin{equation}
I_{n,m}(x)=(-1)^{m}{\frac{{{x^{\frac{{n+m}}{2}}}exp(-x/2)}}{\sqrt{\Gamma
(1+n)\Gamma (1+m)}}}\;,\;%
\mathop{\rm Re}%
x\rightarrow \infty \,\,,  \label{4.30}
\end{equation}
whenever $m$ is integer.

6. One can prove the following asymptotic formula

\begin{equation}
{\lim_{p\rightarrow \infty }}{I_{p+\alpha ,p+\beta }\left( {\frac{x^{2}}{4p}}%
\right) }=J_{\alpha -\beta }(x)\;,  \label{4.40}
\end{equation}
where $J_{\nu }\left( x\right) $ are Bessel functions.\ 

7. Taking into account (\ref{4.29}) and (\ref{4.30}), one can see that only
the functions $I_{\alpha +n,n}(x)$ with non-negative integer $n$ and $\alpha
>-1$ are quadratically integrable on the interval $\left( 0,\infty \right) .$
Such functions obey the orthonormality relation

\begin{equation}
\int\limits_{0}^{\infty }{I_{\alpha +n,n}(x)I_{\alpha +m,m}(x)dx}=\delta
_{m,n}\;,  \label{4.41}
\end{equation}
which follows from the corresponding properties of the Laguerre polynomials
( \cite{GraRy94}, 7.414.3). In such a case, the relation 
\begin{equation}
I_{\alpha +n,n}(x)=\sqrt{{\frac{n!}{{\Gamma (n+\alpha +1)}}}}e^{-{\frac{x}{2}%
}}x^{\frac{\alpha }{2}}L_{n}^{\alpha }(x)\;  \label{4.42}
\end{equation}
follows from (\ref{4.4})

8. Consider a class of functions, which are closely related to Laguerre
functions, and which appear often in various problems of mathematical
physics.

As it follows from Eq. (\ref{4.17}), the Laguerre functions are solutions of
the following eigenvalue problem

\begin{equation}
R_{\alpha }\psi =\lambda \psi ,\;R_{\alpha }=\frac{\alpha ^{2}}{4x}+\frac{x}{%
4}-\frac{d}{dx}-x\frac{d^{2}}{dx^{2}},\quad 0<x<\infty ,\quad \alpha ={\rm %
const}.  \label{4.136}
\end{equation}
A general solution of this problem has the form

\begin{equation}
\psi \left( x\right) =aI_{n,m}\left( x\right) +bI_{m,n}\left( x\right)
,\;\alpha =n-m,\quad 2\lambda =n+m+1,  \label{4.137}
\end{equation}
where $a,b$ are arbitrary constants. In the general case functions $\psi
\left( x\right) $ vanish as $x\rightarrow \infty $ only if one of the
numbers $n$ or $m$ is positive and integer. However, one can provide such a
behavior for any $n,m,$ choosing some special values of $a,b.$ Consider the
functions 
\begin{equation}
\psi _{\lambda ,\alpha }\left( x\right) =x^{-\frac{1}{2}}W_{\lambda ,\frac{%
\alpha }{2}}\left( x\right) ,\quad \psi _{\lambda ,\alpha }\left( x\right)
=\psi _{\lambda ,-\alpha }\left( x\right) ,  \label{4.138}
\end{equation}
where $W_{\lambda ,\mu }\left( x\right) $ are Whittaker functions (\cite
{GraRy94}, 9.220.4). The functions $\psi _{\lambda ,\alpha }\left( x\right) $
can be expressed via the confluent hypergeometric functions 
\begin{eqnarray}
&&\psi _{\lambda ,\alpha }\left( x\right) =e^{-\frac{x}{2}}\left[ \frac{%
\Gamma \left( -\alpha \right) x^{\frac{\alpha }{2}}}{\Gamma \left( \frac{%
1-\alpha }{2}-\lambda \right) }\Phi \left( \frac{1+\alpha }{2}-\lambda
,1+\alpha ;x\right) \right.  \nonumber \\
&&\left. +\frac{\Gamma \left( \alpha \right) x^{-\frac{\alpha }{2}}}{\Gamma
\left( \frac{1+\alpha }{2}-\lambda \right) }\Phi \left( \frac{1-\alpha }{2}%
-\lambda ,1-\alpha ;x\right) \right] ,  \label{4.139}
\end{eqnarray}
or, using (\ref{4.1}), via the Laguerre functions 
\begin{eqnarray}
&&\psi _{\lambda ,\alpha }\left( x\right) =\frac{\sqrt{\Gamma \left(
1+n\right) \Gamma \left( 1+m\right) }}{\sin \left( n-m\right) \pi }\left(
\sin n\pi I_{n,m}\left( x\right) -\sin m\pi I_{m,n}\left( x\right) \right) ,
\nonumber \\
&&\alpha =n-m,\,2\lambda =1+n+m,\,n=\lambda -\frac{1-\alpha }{2},\,m=\lambda
-\frac{1+\alpha }{2}.  \label{4.140}
\end{eqnarray}
By the help of (\ref{4.11})-(4.16),\ the following properties of the
functions $\psi _{\lambda ,\alpha }\left( x\right) $ can be established,

\begin{eqnarray}
\psi _{\lambda ,\alpha }\left( x\right) &=&\sqrt{x}\psi _{\lambda -\frac{1}{2%
},\alpha -1}\left( x\right) +\frac{1+\alpha -2\lambda }{2}\psi _{\lambda
-1,\alpha }\left( x\right) ,  \label{4.141} \\
\psi _{\lambda ,\alpha }\left( x\right) &=&\sqrt{x}\psi _{\lambda -\frac{1}{2%
},\alpha +1}\left( x\right) +\frac{1-\alpha -2\lambda }{2}\psi _{\lambda
-1,\alpha }\left( x\right) ,  \label{4.142} \\
2x\psi _{\lambda ,\alpha }^{\prime }\left( x\right) &=&\left( 2\lambda
-1-x\right) \psi _{\lambda ,\alpha }\left( x\right) +\frac{1}{2}\left(
2\lambda -1-\alpha \right) \left( 2\lambda -1+\alpha \right) \psi _{\lambda
-1,\alpha }\left( x\right) ,  \label{4.143} \\
2x\psi _{\lambda ,\alpha }^{\prime }\left( x\right) &=&\left( \alpha
-x\right) \psi _{\lambda ,\alpha }\left( x\right) +\left( 2\lambda -1-\alpha
\right) \sqrt{x}\psi _{\lambda -\frac{1}{2},\alpha +1}\left( x\right)
\label{4.144} \\
&=&\left( x-2\lambda -1\right) \psi _{\lambda ,\alpha }-2\psi _{\lambda
+1,\alpha }\,.
\end{eqnarray}
As a consequence of these properties we get 
\begin{eqnarray}
&&A_{\alpha }\psi _{\lambda ,\alpha }\left( x\right) =\frac{2\lambda
-1+\alpha }{2}\psi _{\lambda -\frac{1}{2},\alpha -1}\left( x\right)
,\,A_{\alpha }^{+}\psi _{\lambda -\frac{1}{2},\alpha -1}\left( x\right)
=\psi _{\lambda ,\alpha }\left( x\right) ,  \nonumber \\
&&A_{\alpha }=\frac{x+\alpha }{2\sqrt{x}}+\sqrt{x}\frac{d}{dx},\quad
A_{\alpha }^{+}=\frac{x+\alpha -1}{2\sqrt{x}}-\sqrt{x}\frac{d}{dx}.
\label{4.145}
\end{eqnarray}
The operator $R_{\alpha }$ can be expressed via the operators $A_{\alpha
},\;A_{\alpha }^{+},$%
\begin{equation}
R_{\alpha }=A_{\alpha }^{+}A_{\alpha }+\frac{1-\alpha }{2},\quad R_{\alpha
-1}=A_{\alpha }A_{\alpha }^{+}-\frac{\alpha }{2}\;.  \label{4.146}
\end{equation}
Since (\ref{4.140}) is a particular case of (\ref{4.137}), then $\psi
_{\lambda ,\alpha }\left( x\right) $ are also an eigenfunctions for the
operator $R_{\alpha }.$

Using well-known asymptotics of the Whittaker function (\cite{GraRy94},
9.227), we get 
\begin{equation}
\psi _{\lambda ,\alpha }\left( x\right) \sim x^{\lambda -\frac{1}{2}}e^{-%
\frac{x}{2}},\quad x\rightarrow \infty ;\;\;\psi _{\lambda ,\alpha }\left(
x\right) \sim \frac{\Gamma \left( |\alpha |\right) }{\Gamma \left( \frac{%
1+|\alpha |}{2}-\lambda \right) }x^{-\frac{|\alpha |}{2}},\,\,\alpha \neq
0,\quad x\sim 0\,.  \label{4.147}
\end{equation}
The functions $\psi _{\lambda ,0}\left( x\right) $ have a logarithmic
singularity at $x\sim 0.$ It is important to stress that the functions $\psi
_{\lambda ,\alpha }\left( x\right) $ are correctly defined and infinitely
differentiable for $0<x<\infty $ and for any complex $\lambda ,\alpha .$ In
this respect one can mention that the Laguerre functions are not defined for
negative integer $n,m.$ In particular cases, when one of the numbers $n,m$
is non-negative and integer, the functions $\psi _{\lambda ,\alpha }\left(
x\right) $ coincides (up to a constant factor) with Laguerre functions.
Thus, $\psi _{\lambda ,\alpha }\left( x\right) $ are eigenfunctions (of the
operator $R_{\alpha }$), which vanish at $x\rightarrow \infty $ $.$

According to (\ref{4.147}), the functions $\psi _{\lambda ,\alpha }\left(
x\right) $ are quadratically integrable on the interval $0<x<\infty $
whenever $|\alpha |<1$. It is not true for $|\alpha |\geq 1.$ The
corresponding integrals can be calculated (\cite{GraRy94}, 7.611),

\begin{eqnarray}
&&\int\limits_{0}^{\infty }\psi _{\lambda ,\alpha }\left( x\right) \psi
_{\lambda ^{\prime },\alpha }\left( x\right) \;dx=\frac{\pi }{\left( \lambda
^{\prime }-\lambda \right) \sin \alpha \pi }\left\{ \left[ \Gamma \left( 
\frac{1+\alpha -2\lambda ^{\prime }}{2}\right) \Gamma \left( \frac{1-\alpha
-2\lambda }{2}\right) \right] ^{-1}\right.  \nonumber \\
&&\left. -\left[ \Gamma \left( \frac{1-\alpha -2\lambda ^{\prime }}{2}%
\right) \Gamma \left( \frac{1+\alpha -2\lambda }{2}\right) \right]
^{-1}\right\} ,\quad |\alpha |<1,  \label{4.148} \\
&&\int\limits_{0}^{\infty }|\psi _{\lambda ,\alpha }\left( x\right)
|^{2}\;dx=\frac{\pi }{\sin \alpha \pi }\frac{\psi \left( \frac{1+\alpha
-2\lambda }{2}\right) -\psi \left( \frac{1-\alpha -2\lambda }{2}\right) }{%
\Gamma \left( \frac{1+\alpha -2\lambda }{2}\right) \Gamma \left( \frac{%
1-\alpha -2\lambda }{2}\right) },\quad |\alpha |<1,  \label{4.149} \\
&&\int\limits_{0}^{\infty }|\psi _{\lambda ,0}\left( x\right) |^{2}\;dx=%
\frac{\psi ^{\prime }\left( \frac{1}{2}-\lambda \right) }{\Gamma ^{2}\left( 
\frac{1}{2}-\lambda \right) },\quad \int\limits_{0}^{\infty }\left| \psi _{n+%
\frac{1}{2},0}\left( x\right) \right| ^{2}dx=\Gamma ^{2}\left( 1+n\right) .
\label{4.150}
\end{eqnarray}
Here $\psi (x)$ is the logarithmic derivative of the $\Gamma -$function ( 
\cite{GraRy94}, 8.360).

For $|\alpha |\geq 1,$ the situation is the following: the only
quadratically integrable eigenfunctions of the operator $R_{\alpha }$ are
Laguerre functions, they also form a complete set. The functions $\psi
_{\lambda ,\alpha }\left( x\right) $ are orthogonal whenever arguments of
the $\Gamma -$function in (\ref{4.148}) are integer and negative. That
corresponds to $n,m$ integer and non-negative. Thus, that is again the case
of Laguerre functions according to (\ref{4.140}). If $|\alpha |<1,$ then, in
the general case, the functions $\psi _{\lambda ,\alpha }\left( x\right) \;$%
and $\psi _{\lambda ^{\prime },\alpha }(x),$ $\lambda ^{\prime }\neq \lambda
,$ are not orthogonal, as it follows from (\ref{4.148}). That is a
reflection of the fact that $R_{\alpha }$ is not anymore self-conjugate
operator for such values of $\alpha .$

\end{document}